\documentclass[11pt]{amsart}
\usepackage{geometry}                
\usepackage{graphicx}
\usepackage{amssymb}
\usepackage{epstopdf}
\usepackage{hyperref}
\usepackage{array}
\usepackage{float}

\newcolumntype{L}[1]{>{\raggedright\let\newline\\\arraybackslash\hspace{0pt}\vspace{0pt}}m{#1}}
\newcolumntype{C}[1]{>{\centering\let\newline\\\arraybackslash\hspace{0pt}}m{#1}}
\newcolumntype{R}[1]{>{\raggedleft\let\newline\\\arraybackslash\hspace{0pt}\vspace{0pt}}m{#1}}
\title{An interdisciplinary survey of network similarity methods}
\author{Emily Evans and Marissa Graham}

\begin{document}
\begin{abstract}Comparative graph and network analysis plays an important role in both systems biology and pattern recognition, but existing surveys on the topic have historically ignored or underserved one or the other of these fields. We present an integrative introduction to the key objectives and methods of graph and network comparison in each field, with the intent of remaining accessible to relative novices in order to mitigate the barrier to interdisciplinary idea crossover. 

To guide our investigation, and to quantitatively justify our assertions about what the key objectives and methods of each field are, we have constructed a citation network containing 5,793 vertices from the full reference lists of over two hundred relevant papers, which we collected by searching Google Scholar for ten different network comparison-related search terms. We investigate its basic statistics and community structure, and frame our presentation around the papers found to have high importance according to five different standard centrality measures.\end{abstract}
\maketitle

Quantitatively comparing objects and systems is a common question across scientific disciplines. Networks are a powerful data structure for the representation of complex objects and systems, but the same generality that makes them powerful also makes them computationally difficult to work with. Over the past twenty years or so, however, the question of quantitative network comparison has gained significant attention as computational resources and interesting network data have become more readily available. Tools for analyzing this network data are relevant to mathematicians, biologists, computer and social scientists alike.

When discussing the similarity of networks, we can compare the vertices in a single network to each other, or we can compare entire networks. In this work, we focus on the latter question: both the direct comparison of two networks, and the construction of statistics that allow us to classify networks according to certain meaningful properties.

Our approach differs from a standard survey paper; in addition to studying the tools of network theory, we used them to guide our reading. We have constructed a network of citations between scientific papers on this topic, which allows us to use standard network analysis techniques to determine which papers are the most important or influential and to more easily consider the context of the field as a whole. We can therefore quantitatively justify our assertions using standard centrality and community detection measures instead of relying on existing expertise in the field to give weight to our claims. 

The rest of this work is structured as follows. In Section~\ref{sec:dataset_creation_and_analysis}, we introduce our citation network dataset, analyze its basic structure, and discuss the two main fields of application within the study of network comparison. In Section~\ref{sec:centrality}, we introduce the centrality measures used in our analysis and provide a list of the high centrality vertices in our dataset, i.e., the papers which are most important or influential. In Sections \ref{chapter:pattern_recognition} and \ref{chapter:systems_biology}, we discuss commonly used network similarity techniques from the fields of pattern recognition and biology, respectively. We then conclude in Section~\ref{sec:conclusion} by discussing potential cross-applications between these two fields.

\section{Dataset Creation and Statistics}\label{sec:dataset_creation_and_analysis}

Our first task was to create a citation network to guide our reading.
While some journals and databases provide citation networks of the references in their own domain, investigating only the citations within a single database or journal would not have given us the interdisciplinary picture we were looking for. As a result, and since intellectual property restrictions preclude directly scraping an entire citation network, we manually constructed the dataset from the reference lists of relevant papers.

Relevant papers were found by searching Google Scholar for ``graph" or ``network" +  ``alignment", ``comparison", ``similarity", ``isomorphism", or ``matching". We initially collected topic-relevant papers from the first five pages of results for each of these ten search terms on May 4th, 2018, after which we collected new papers published through June 25th, 2018 from a Google Scholar email alert for the same ten search terms. We stored the plaintext reference list or each paper in a standardized format that could be easily split into the individual freeform citations. We refer to papers for which we have a reference list as parents, and to their references as children. In total, we collected 7,790 child references from 221 parent papers.

To create a network from these reference lists, we collected the metadata of each freeform citation in order to disambiguate the same work across different freeform citations.
Our first step was to search for each citation in the CrossRef database using the REST API \cite{crossrefAPI}, which returns parsed metadata for the (possibly incorrect) best match it finds. We considered two citations to refer to the same paper if their metadata matched and was known to be correct for both, or if both their metadata and freeform citations matched exactly. 

We considered the results of the REST API to be correct if the title it returned was present in the original freeform citation and unverified otherwise. We then manually handled the unverified records. For unverified parents, we found the title, year, author, DOI, and URL information, as well as reference and citation counts. We then checked the unverified child references. About half of these were correct, but were not automatically verified due to punctuation discrepancies, misspellings, Unicode issues, or citation styles that do not include a title. Some results consisted of a review, purchase listing, or similar for the correct record; we marked those as correct, but noted them as ``half-right". For the remaining incorrect references, we manually added the author, title, and year from the citation. Finally, we deleted any records which did not refer to a written work of some kind, i.e, those citing a website, web service, database, software package/library, programming language, or personal communication.

The dataset itself and the code and source files used to generate it can be found at \url{https://github.com/marissa-graham/network-similarity}, along with documentation and instructions for using it to generate a similar dataset for any collection of properly-formatted reference list files.
\subsection{Basic statistics}

Our full citation network contains 7,491 references between 5,793 papers. This is an abnormally small fraction of vertices with children compared to a typical citation network such as the Garfield networks included in Table \ref{tab:network_table}, but this is a result of our construction method. Unlike our network, the Garfield datasets each contain references from a single database; they have child information for more references, but references outside the database are pruned.

We also consider the \textit{pruned network}, which we define to be the giant component of the subnetwork of vertices with either a positive outdegree or an indegree greater than one. That is, we discard papers which are not part of the giant component as well as child papers which are not cited by multiple parents, since these are less likely to be relevant to the field of network similarity. Doing this decreases the complexity of the network while preserving its most relevant vertices. This restriction shrinks the number of vertices by a factor of almost six, correspondingly
raises the fraction of vertices with children, and approximately doubles the mean degree.

In Table \ref{tab:network_table}, we display the mean degree, fraction of vertices with children, diameter, number of connected components, and the fraction of vertices in the giant component for five different networks: our full network $G$, its pruned version $G_p$, two datasets from the Garfield citation network collection, and a uniformly generated random graph $R$ of the same size as $G$. We notice increased connectivity in $G$ and $G_p$ compared to both the Garfield datasets and to a random control.

\begin{table}[t]
\centering
\begin{tabular}{|l|c|c|c|c|c|}
\hline
 & $G$ & $G_p$ & SciMet & Zewail & $R$ \\ \hline\hline
Vertices & 5793 & 1062 & 1092 & 3145 & 1077 \\ \hline 
Edges & 7491 & 2775 & 1308 & 3743 & 7491\\ \hline 
Mean degree & 1.29 & 2.61 & 1.20 & 1.19 & 1.29\\ \hline
Fraction with children & 0.038 & 0.193 & 0.523 & 0.599 & 0.733 \\ \hline
Diameter & 10 & 9 & 14 & 22 & 21 \\ \hline 
Connected components & 16 & 1 & 114 & 281 & 504 \\ \hline
Fraction in giant component & 0.960 & 1.000 & 0.784 & 0.797 & 0.900  \\ \hline 
\end{tabular}
\vspace{0.2cm}
\caption{Comparing statistics for our dataset to other networks.}

\label{tab:network_table}
\end{table}

\label{section:partitioning}
\subsection{Network Partitioning}

\begin{figure}[h]
\centering
\includegraphics[width=0.7\textwidth]{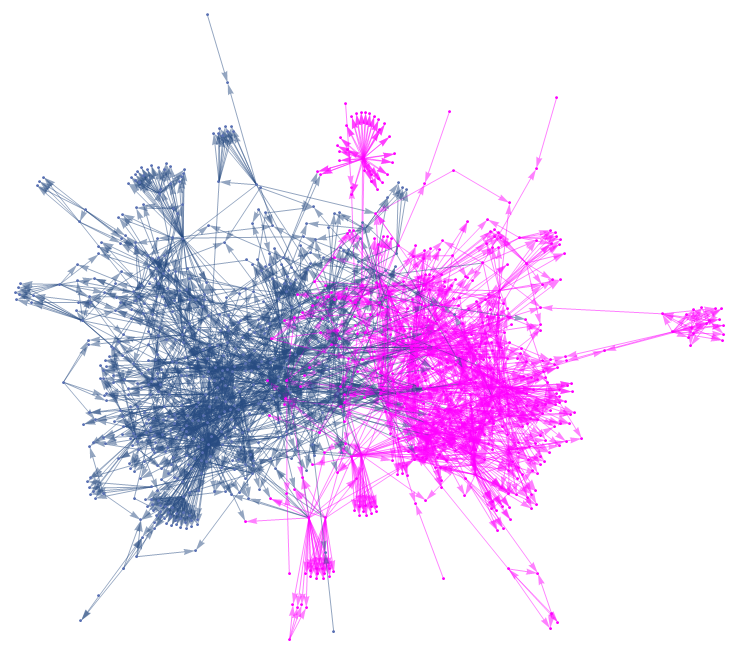}
\caption{The two halves of our partition of the pruned network.}
\label{fig:partitioned_network}
\end{figure}

The applications of both the parent papers in our citation network and the unverified child references we manually checked were found almost exclusively in the fields of biology and computer science, and we suspected that this was reflected in the structure of the network. Unfortunately, we did not have subject metadata that would have allowed us to partition the network with respect to these categories. Instead, we performed a modularity maximizing partition to separate the network into the two clusters of more densely connected vertices shown in Figure \ref{fig:partitioned_network}, and then investigated whether those clusters corresponded to biology and computer science. 

We did so by tagging the titles of the journals each paper was published in according to the keywords listed in Table \ref{tab:tagging_keywords}. This strategy gave us subject information for about 67\% of the total papers and 53\% of those in the pruned network, which we found sufficient to confirm our initial suspicions that our structural partition of the network reflects the two fields of application we observed while constructing it.
\begin{figure}[p]
\centering
\begin{minipage}[c]{0.23\textwidth}
\includegraphics[width=\textwidth]{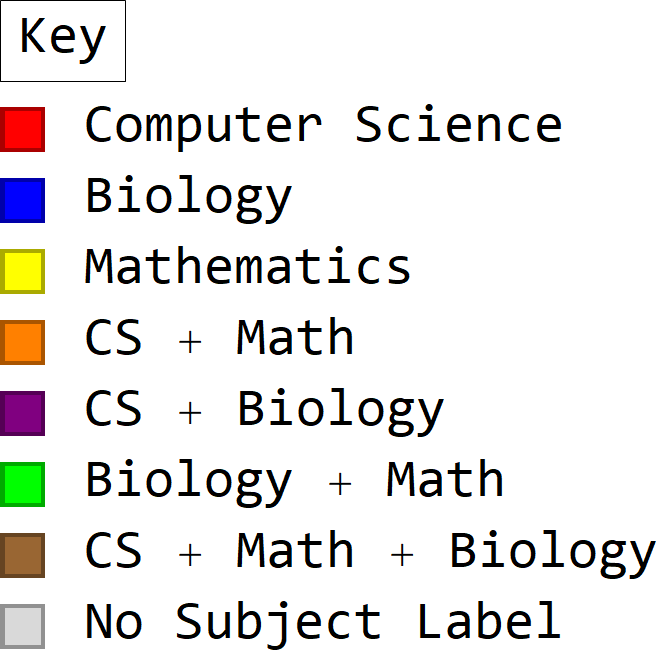}
\end{minipage}
\hfill
\begin{minipage}[c]{0.7\textwidth}
a)\includegraphics[width=0.95\textwidth]{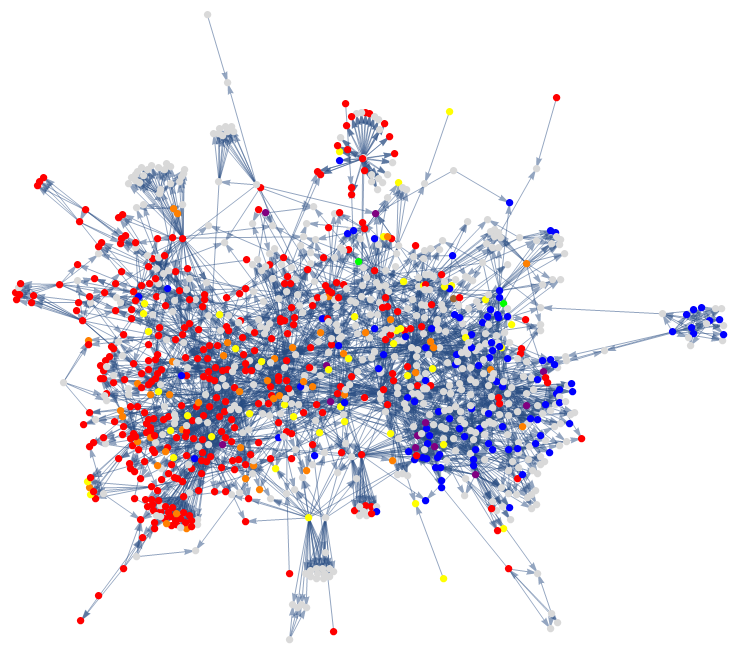}
\end{minipage}
\begin{minipage}[c]{0.49\textwidth}
\includegraphics[width=0.95\textwidth]{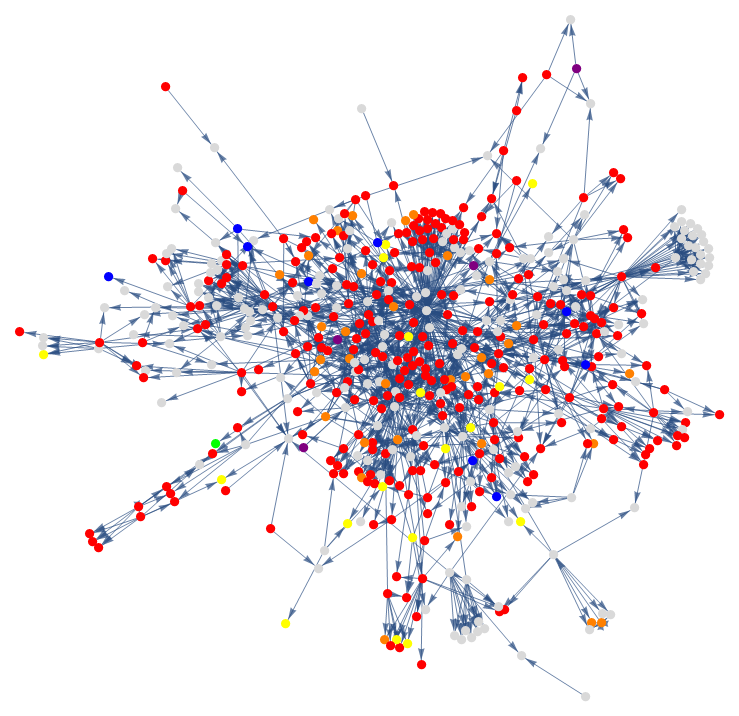}

b)
\vspace{-16pt}
\end{minipage}
\hfill
\begin{minipage}[c]{0.49\textwidth}
\includegraphics[width=0.95\textwidth]{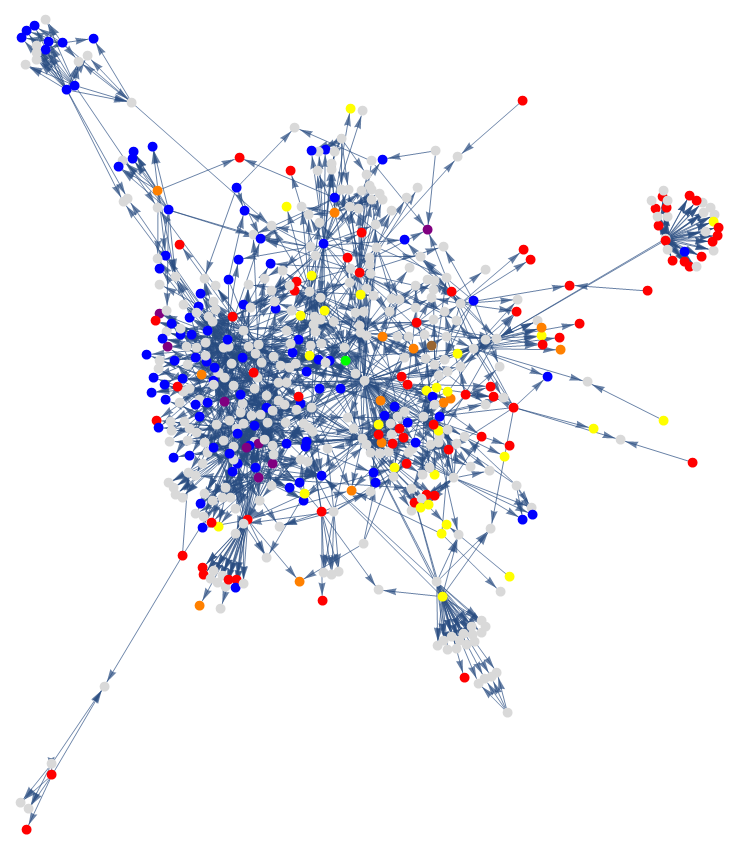}
\end{minipage}
\caption{a) The pruned network $G_p$, and b) the two halves of its partition $G_1$ and $G_2$, with vertices colored according to their subject label.}
\label{fig:subject_color_coded}
\end{figure}
%
%
%
%
%
%
%
%

While we are primarily interested in computer science and biology, we tagged mathematics papers as well to obtain a third category of similar size and generality to the other two. It serves as a control group. We then counted the vertices with each tag on either side of our partition, as shown in Table \ref{tab:subject_counts}. There are 312 CS papers in $G_2$ but only 93 in $G_1$, and there are 108 biology papers in $G_1$ but only 14 in $G_2$, but the math category is fairly evenly distributed across the two halves of the partition, with 44 papers in $G_1$ and 53 in $G_2$. We therefore conclude that the two halves of our structurally constructed partition approximately correspond to the two fields of application we expected it would.

We note that there are significantly more untagged vertices on the biology side of the partition. This is likely because the computer science category is both inherently larger and more likely to be tagged as such; the majority of computer science papers are published in an ACM, IEEE, or SIAM journal\footnote{Note that ``SIAM" and ``algorithm" were used as keywords for both math and computer science, which accounts for about half of the overlap between the two categories.}, but there does not seem to be an analogous group of dominant acronymic organizations in biology to allow for easy tagging of journal titles.

\begin{table}[t]
\centering
\setlength\extrarowheight{3pt}
\begin{tabular}{|l|r|r|r|r|}
\hline & $G$ & $G_p$ & $G_1$ & $G_2$ \\ \hline\hline
Total vertices & 5793 & 1062 & 531 & 531 \\ \hline
Untagged & 1922 & 502 & 311 & 191 \\ \hline
Tagged & 3871 & 560 & 220 & 340 \\ \hline
CS & 2533 & 405 & 93 & 312 \\ \hline
Biology & 984 & 122 & 108 & 14 \\ \hline
Math & 787 & 97 & 44 & 53 \\ \hline
Both CS and biology & 108 & 13 & 9 & 4 \\ \hline
Both CS and math & 305 & 49 & 15 & 34 \\ \hline
Both biology and math & 24 & 3 & 2 & 1 \\ \hline
All three & 4 & 1 & 1 & 0 \\ \hline
\end{tabular}
\vspace{0.2cm}
\caption{Number of vertices tagged as computer science, biology, math, or some combination of these in $G$, $G_p$, and the two halves of the partition $G_1$ and $G_2$.}
\label{tab:subject_counts}
\end{table}

\section{Centrality}\label{sec:centrality}

Many different centrality measures are used in network theory, not all of which are relevant to our network and for our purposes. We have chosen the following five to drive our choice of references to discuss in this work:
\begin{itemize}
\item indegree (the number of times each paper was cited),
\item outdegree (the number of papers cited within the pruned network),
\item betweenness (the extent which a vertex lies on paths between other vertices),
\item closeness\footnote{Recall that the closeness centrality is determined by taking the inverse of the average geodesic distance between a vertex and all other vertices in the network. Closeness centrality therefore takes on the highest values for a vertex which has a short average distance from all other vertices.}, and 
\item HITS (hyperlink-induced topic search).
\end{itemize} 

Clearly papers which are frequently cited tend to be important, but papers with high outdegree tend to be relevant surveys and are therefore worth considering. Papers with high betweenness centrality tend to make connections between disparate ideas in an original way, and papers with high closeness centrality tend to be relevant to the field as a whole. 

The HITS algorithm allows us to separately consider the notion that a vertex is important if it points to other important vertices, and the notion that a vertex is important if it is pointed to by other important vertices. It defines two different types of centrality for each vertex. The \emph{authority centrality}\index{authority centrality} of a vertex measures whether it is being pointed to by vertices with high \emph{hub centrality}\index{hub centrality}, which in turn measures whether those vertices point to vertices with high authority centrality. By defining the hub and authority centralities of a vertex to be proportional to the sum of the authority and hub centralities, respectively, of its neighbors, this definition reduces to a pair of eigenvalue equations which can be easily solved numerically. 

We consider the most central papers for both the entire network and for each half of the partition discussed in Section \ref{section:partitioning}. For each of these, we collected the top ten papers according to our six centrality metrics\footnote{We have discussed five centrality measures, but since HITS defines two centrality types, we have six centrality metrics.} and ranked their relevance with respect to \[ f(p) = \frac{k^{in}_p}{k^{in}_{max}} + \frac{k^{out}_p}{ k^{out}_{max}} + \sum_{i=1}^4 \frac{1}{r_i(p)}, \] where $k^{in}_p$ and $k^{out}_p$ are the indegree and outdegree of a paper $p$, the maximums of which are taken with respect to the pruned network or partition half in question, and $r_i(p)$ is the rank of a paper $p$ according to the $i$-th of our four centrality metrics, which is defined to be infinity if a paper is not in the top ten for that metric. 

Tables \ref{tab:toppapers_all}, \ref{tab:toppapers_bio}, and \ref{tab:toppapers_CS} show the top twenty papers in the entire network and each side of the partition with respect to $f$. Since the numerical values for indegree and outdegree have intuitive meaning, we report the value itself. However, the values for betweenness, closeness, and the two HITS centralities are unintuitive, context-free real numbers, so we report the rank of each paper with respect to each of these measures rather than the actual value. We also calculate betweenness and closeness for the undirected version of the network to allow those rankings to be based on citing relationships in either direction.

\begin{table}[h]

\centering
{\setlength\extrarowheight{2pt}\setlength{\tabcolsep}{3pt}
\begin{tabular}{|p{10cm}|c|c|c|c|c|c|}
\hline
& \rotatebox[origin=c]{90}{Indegree} &  \rotatebox[origin=c]{90}{Outdegree} & \rotatebox[origin=c]{90}{Betweenness} &  \rotatebox[origin=c]{90}{Closeness} &  \rotatebox[origin=c]{90}{HITS Auth.} & \rotatebox[origin=c]{90}{HITS Hub} \\ 
\hline\hline
$^\Diamond$Thirty years of graph matching in pattern recognition  \cite{Conte_2004} & 20* & 109* & 1 & 2 &  & 1 \\ \hline
$\dagger$Fifty years of graph matching, network alignment and network comparison  \cite{Emmert_Streib_2016} & 6 & 71* & 2 & 1 &  & 3 \\ \hline
$\dagger$Networks for systems biology: Conceptual connection of data and function  \cite{Emmert_Streib_2011} & 2 & 102* & 3 & 3 &  & 2 \\ \hline
$^\Diamond$An algorithm for subgraph isomorphism  \cite{Ullmann_1976} & 20* & 4 & 7 & 4 & 1 &  \\ \hline
$\dagger$Modeling cellular machinery through biological network comparison  \cite{Sharan_2006} & 9 & 41* & 8 &  &  &  \\ \hline
$^\Diamond$Computers and intractability: A guide to the theory of NP-completeness  \cite{Hartmanis_1982} & 16* & 0 & 4 & 5 &  &  \\ \hline
$^\Diamond$The graph matching problem  \cite{Livi_2012} & 2 & 55* & 5 & 6 &  & 7 \\ \hline
$\dagger$A new graph-based method for pairwise global network alignment  \cite{Klau_2009} & 9 & 13 &  & 8 &  &  \\ \hline
$\dagger$On graph kernels: Hardness results and efficient alternatives  \cite{Gartner_2003} & 11 & 10 & 6 &  &  &  \\ \hline
$^\Diamond$Error correcting graph matching: On the influence of the underlying cost function  \cite{Bunke_1999} & 10 & 16 &  & 7 & 7 & 8 \\ \hline
$^\Diamond$A graduated assignment algorithm for graph matching  \cite{Gold_1996} & 18* & 0 &  &  & 5 &  \\ \hline
$^\Diamond$The Hungarian method for the assignment problem  \cite{Kuhn_1955} & 17* & 0 &  &  &  &  \\ \hline
$^\Diamond$An eigendecomposition approach to weighted graph matching problems  \cite{Umeyama_1988} & 15* & 5 &  &  & 6 &  \\ \hline
$^\Diamond$Recent developments in graph matching  \cite{Bunke_2000} & 1 & 51* &  &  &  & 4 \\ \hline
$\dagger$MAGNA: Maximizing accuracy in global network alignment  \cite{Saraph_2014} & 5 & 35* &  &  &  &  \\ \hline
$^\Diamond$A distance measure between attributed relational graphs for pattern recognition  \cite{Sanfeliu_1983} & 14* & 0 &  &  & 3 &  \\ \hline
$\dagger$Pairwise global alignment of protein interaction networks by matching neighborhood topology  \cite{Singh_2007} & 13* & 0 &  &  &  &  \\ \hline
$\dagger$Topological network alignment uncovers biological function and phylogeny  \cite{Bunke_1998} & 12* & 0 &  &  &  &  \\ \hline
A graph distance metric based on the maximal common subgraph  \cite{Kuchaiev_2010} & 10 & 0 &  & 10 & 4 &  \\ \hline
$^\Diamond$Efficient graph matching algorithms  \cite{Messmer_1995} & 0 & 43* &  &  &  & 5 \\ \hline
\end{tabular}

$\dagger$Also top for Group 1 (biology dominated); $^\Diamond$Also top for Group 2 (computer science dominated)
}
\caption{Highest centrality papers for the entire pruned network.}
\label{tab:toppapers_all}
\end{table}

\begin{table}[h]
\centering
{\setlength\extrarowheight{2pt}\setlength{\tabcolsep}{3pt}
\begin{tabular}{|p{10cm}|c|c|c|c|c|c|}
\hline
& \rotatebox[origin=c]{90}{Indegree} &  \rotatebox[origin=c]{90}{Outdegree} & \rotatebox[origin=c]{90}{Betweenness} &  \rotatebox[origin=c]{90}{Closeness} &  \rotatebox[origin=c]{90}{HITS Auth.} & \rotatebox[origin=c]{90}{HITS Hub} \\ 
\hline\hline
$^\Diamond$Networks for systems biology: Conceptual connection of data and function  \cite{Emmert_Streib_2011} & 2 & 90* & 1 & 2 &  & 1 \\ \hline
$^\Diamond$Fifty years of graph matching, network alignment and network comparison  \cite{Emmert_Streib_2016} & 4 & 56* & 2 & 1 &  & 2 \\ \hline
$^\Diamond$Modeling cellular machinery through biological network comparison  \cite{Sharan_2006} & 9 & 40* & 4 & 3 & 10 & 9 \\ \hline
$^\Diamond$MAGNA: Maximizing accuracy in global network alignment  \cite{Saraph_2014} & 5 & 35* & 7 & 6 &  & 3 \\ \hline
$^\Diamond$On graph kernels: Hardness results and efficient alternatives  \cite{Gartner_2003} & 10* & 9 & 3 & 8 &  &  \\ \hline
Biological network comparison using graphlet degree distribution  \cite{Przulj_2007} & 11* & 0 &  & 7 & 4 & 7 \\ \hline
$^\Diamond$A new graph-based method for pairwise global network alignment  \cite{Klau_2009} & 8 & 12 & 9 & 4 & 6 &  \\ \hline
Network motifs: Simple building blocks of complex networks  \cite{Milo_2002} & 11* & 0 &  & 9 & 8 &  \\ \hline
$^\Diamond$Pairwise global alignment of protein interaction networks by matching neighborhood topology  \cite{Singh_2007} & 12* & 0 &  &  & 3 &  \\ \hline
$^\Diamond$Topological network alignment uncovers biological function and phylogeny  \cite{Kuchaiev_2010} & 12* & 0 &  &  & 2 &  \\ \hline
NETAL: A new graph-based method for global alignment of protein-protein interaction networks  \cite{Neyshabur_2013} & 6 & 26* &  &  &  & 5 \\ \hline
Collective dynamics of ``small-world" networks  \cite{Watts_1998} & 10* & 0 &  & 10 & 5 &  \\ \hline
Global network alignment using multiscale spectral signatures  \cite{Patro_2012} & 11* & 0 &  &  & 9 &  \\ \hline
$^\Diamond$Global alignment of multiple protein interaction networks with application to functional orthology detection  \cite{Singh_2008} & 10* & 0 &  &  &  &  \\ \hline
Conserved patterns of protein interaction in multiple species  \cite{Sharan_2005} & 10* & 0 &  &  & 7 &  \\ \hline
Pairwise alignment of protein interaction networks  \cite{Koyuturk_2006} & 10* & 0 &  &  & 1 &  \\ \hline
Alignment-free protein interaction network comparison  \cite{Ali_2014} & 2 & 22 & 6 & 5 &  &  \\ \hline
Graphlet-based measures are suitable for biological network comparison  \cite{Hayes_2013} & 1 & 30* &  &  &  & 8 \\ \hline
Survey on the graph alignment problem and a benchmark of suitable algorithms  \cite{Dopmann_2013} & 0 & 26 &  &  &  & 4 \\ \hline
Predicting graph categories from structural properties  \cite{Canning_2018} & 0 & 30* & 5 &  &  &  \\ \hline
\end{tabular}

$^\Diamond$Also a top-centrality paper for the entire network}
\centering
\caption{Highest centrality papers for the biology-dominated half of the pruned network.}
\label{tab:toppapers_bio}
\end{table}

\begin{table}[h]
\centering
{\setlength\extrarowheight{2pt}\setlength{\tabcolsep}{3pt}
\begin{tabular}{|p{10cm}|c|c|c|c|c|c|}
\hline
& \rotatebox[origin=c]{90}{Indegree} &  \rotatebox[origin=c]{90}{Outdegree} & \rotatebox[origin=c]{90}{Betweenness} &  \rotatebox[origin=c]{90}{Closeness} &  \rotatebox[origin=c]{90}{HITS Auth.} & \rotatebox[origin=c]{90}{HITS Hub} \\ 
\hline\hline
$^\Diamond$Thirty years of graph matching in pattern recognition  \cite{Conte_2004} & 17* & 107* & 1 & 1 &  & 1 \\ \hline
$^\Diamond$An algorithm for subgraph isomorphism  \cite{Ullmann_1976} & 15* & 2 & 10 & 5 & 2 &  \\ \hline
$^\Diamond$A graduated assignment algorithm for graph matching  \cite{Gold_1996} & 18* & 0 & 7 & 4 & 3 &  \\ \hline
$^\Diamond$An eigendecomposition approach to weighted graph matching problems  \cite{Umeyama_1988} & 15* & 5 &  & 2 & 4 &  \\ \hline
$^\Diamond$The graph matching problem  \cite{Livi_2012} & 2 & 36* & 3 & 3 &  & 8 \\ \hline
$^\Diamond$A distance measure between attributed relational graphs for pattern recognition  \cite{Sanfeliu_1983} & 13* & 0 &  & 7 & 1 &  \\ \hline
$^\Diamond$Recent developments in graph matching  \cite{Bunke_2000} & 0 & 50* & 8 &  &  & 2 \\ \hline
$^\Diamond$Error correcting graph matching: On the influence of the underlying cost function  \cite{Bunke_1999} & 9* & 16 &  & 8 &  & 6 \\ \hline
$^\Diamond$Fast and scalable approximate spectral matching for higher order graph matching  \cite{Park_2014} & 0 & 41* & 2 &  &  &  \\ \hline
$^\Diamond$Efficient graph matching algorithms  \cite{Messmer_1995} & 0 & 42* & 5 &  &  & 4 \\ \hline
$^\Diamond$Computers and intractability: A guide to the theory of NP-completeness  \cite{Hartmanis_1982} & 11* & 0 & 6 &  &  &  \\ \hline
$^\Diamond$The Hungarian method for the assignment problem  \cite{Kuhn_1955} & 14* & 0 &  &  &  &  \\ \hline
$^\Diamond$Graph matching applications in pattern recognition and image processing  \cite{Conte_2003} & 0 & 40* &  &  &  & 3 \\ \hline
Efficient graph similarity search over large graph databases  \cite{Zheng_2015} & 0 & 28* & 4 & 6 &  &  \\ \hline
A linear programming approach for the weighted graph matching problem  \cite{Almohamad_1993} & 8 & 8 &  & 9 & 9 &  \\ \hline
$^\Diamond$Structural matching in computer vision using probabilistic relaxation  \cite{Christmas_1995} & 9* & 0 &  &  & 5 &  \\ \hline
$^\Diamond$A graph distance measure for image analysis  \cite{Eshera_1984} & 8 & 0 &  &  & 6 &  \\ \hline
Inexact graph matching for structural pattern recognition  \cite{Bunke_1983} & 10* & 0 &  &  &  &  \\ \hline
$^\Diamond$A new algorithm for subgraph optimal isomorphism  \cite{El_Sonbaty_1998} & 2 & 21 &  &  &  & 5 \\ \hline
Approximate graph edit distance computation by means of bipartite graph matching  \cite{Riesen_2009} & 9 & 0 &  &  &  &  \\ \hline
\end{tabular}

$^\Diamond$Also a top-centrality paper for the entire network}
\centering
\caption{Highest centrality papers for the CS-dominated half of the pruned network.}
\label{tab:toppapers_CS}
\end{table}

The combination of the rankings in these tables, our other available metadata, and the network itself gives us a powerful amount of context to guide our discussion. We can compare an author's framing of their references to our own high centrality papers, follow the forward references of papers with high authority centrality to track the development of the field over time, check cocitations to investigate the idea transfer between disciplines, and so on, which we found to be a significant advantage for our presentation of the topics discussed in the remainder of this work.

\newpage
\section{Pattern Recognition}\label{chapter:pattern_recognition}
 
The combinatorial nature of graphs makes them computationally difficult to work with, but it also makes them a powerful data structure that can be adapted to represent various objects and concepts. Graphs are particularly useful in computer vision, where they can overcome differences at the pixel level as a result of things like angles, lighting, and image scaling. Since graphs are invariant under positional changes including translations, rotations, and mirroring, they are well suited for this task.

Applications in the area of computer vision include optical character recognition \cite{Lu_1991,Rocha_1994}, biometric identification \cite{isenor1986fingerprint,deng2010retinal} and medical diagnostics \cite{sharma2012determining}, and 3D object recognition \cite{Christmas_1995}. As a sample in the past year graphs have been used to recognize Indian sign language~\cite{Kumar_2018a,Kumar_2018b}, spot subgraphs (e.g., certain characters) in comic book images \cite{le2018ssgci}, and to stack MRI image slices \cite{clough2018mri}. A more comprehensive timeline can be found in~\cite{Conte_2004}.

Once we have a graph representation of objects we would like to compare, the problem of object recognition--and in particular, the problem of a database search--is reduced to a graph matching problem. 
\subsection{The graph matching problem}\label{section:defining_graph_matching}

In the literature, the term ``graph matching"\index{graph matching} is frequently used  without being explicitly defined. In cases where a definition is given, it is usually tailored to the purposes of the paper and specific to a certain type of graph matching; i.e., exact, inexact, error-correcting, bipartite, and so on. The distinctions between these can be subtle; for example, the question of finding a matching \emph{in} a graph \cite{wikiMatchingInAGraph} is related to but distinct from graph matching, where we find a matching \emph{between} two graphs, and there is a significant presence in the literature of \emph{elastic graph matching}, which is common in pattern recognition but is not in fact a form of graph matching \cite{Conte_2003}.  In this section, we give an overview of graph matching-related terms and summarize their distinctions.

\subsubsection{Variants of the graph isomorphism problem}
Graph isomorphism is the strictest form of graph matching. A \emph{graph isomorphism}\index{graph isomorphism} is a bijective mapping between the nodes of two graphs of the same size which is \emph{edge-preserving}\index{edge-preserving}; that is, if two nodes in the first graph are connected by an edge, they are mapped to two nodes in the second graph which are also connected by an edge \cite{Conte_2004}.  Likewise if two nodes are not connected, this unconnectedness is also preserved. The decision problem of determining whether two graphs are isomorphic is neither known to be in NP nor known to be solvable in polynomial time \cite{wikiGraphIsomorphism,Hartmanis_1982}.

An \emph{induced subgraph}\index{induced subgraph} of a graph is a graph formed from a subset of nodes in the larger graph and all the edges between them. By contrast, a \emph{subgraph}\index{subgraph} is simply a graph formed from a subset of the nodes and edges in the larger graph.
\begin{figure}[!t]
\centering
\includegraphics[width=\textwidth]{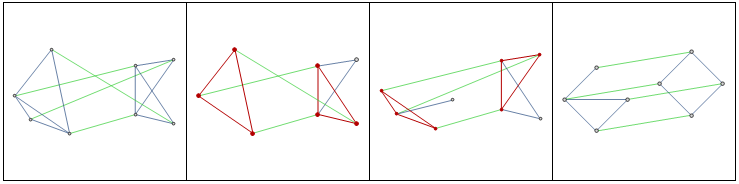}
\caption{A visual summary of the distinctions between (from left to right) graph isomorphism, subgraph isomorphism, maximum common subgraph, and inexact matching.}
\label{fig:isomorphism_demos}
\end{figure}

A \emph{subgraph isomorphism}\index{subgraph isomorphism} is an edge-preserving injective mapping from the nodes of a smaller graph to the nodes of a larger graph. That is, there is a subgraph isomorphism between two graphs if there is an isomorphism between the smaller graph and some subgraph of the larger~\cite{Conte_2004}. The decision problem of determining whether a graph contains a subgraph which is isomorphic to some smaller graph is known to be NP-complete~\cite{wikiSubgraphIsomorphism}.
Finally, a \emph{maximum common induced subgraph}\index{maximum common subgraph} (MCS) of two graphs is a graph which is an induced subgraph of both and has as many nodes as possible  \cite{wikiMaximumCommonSubgraph}. We can formulate the MCS problem as a graph matching problem using the metric \[d(G_1,G_2) = 1 - \frac{|\text{MCS}(G_1,G_2)|}{\max\{|G_1|,|G_2|\}},\] where $|G|$ is the number of nodes in the graph~\cite{Bunke_1998,Bunke_1997}.
\subsubsection{Exact and inexact matching}\label{section:exact_and_inexact_matching}

%
A graph matching method is \emph{exact}\index{exact matching} if it seeks to find a mapping between the nodes of two graphs which is edge-preserving; that is, if two nodes in the first graph are connected by an edge, they are mapped to two nodes in the second graph which are also connected by an edge \cite{Conte_2004}. Exact matching is also sometimes defined by whether a method seeks a boolean evaluation of the similarity of two graphs \cite{Livi_2012,Emmert_Streib_2016}. For graph and subgraph isomorphism, this characterization is equivalent; either two graphs are isomorphic, or they are not. Since the maximum common subgraph problem is edge preserving, we consider it in this work to be an exact matching problem. However, it does not seek a boolean evaluation, and it is therefore sometimes considered to be an inexact matching problem \cite{Livi_2012}. 

By contrast, an \emph{inexact} graph matching method is not edge-preserving. Inexact matching allows us to compensate for the inherent variability of data as well as the variability introduced by constructing a graph representation of data. Instead of forbidding matchings between nodes if edge-preservation requirements are unsatisfied, a penalty is applied, and we seek to minimize the sum of this penalty over all matched nodes. Instead of returning a value in $\{0,1\}$, we return a value in $[0,1]$ which quantifies the similarity or dissimilarity between two graphs. 
Inexact matching algorithms which are based on an explicit cost function or edit distance are often called \emph{error tolerant}\index{error tolerant} or \emph{error correcting}\index{error correcting}. %

\subsubsection{Optimal and approximate algorithms}

Problem formulations for inexact matching generally seek to minimize some nonnegative cost function which should theoretically be zero for two isomorphic graphs. An \emph{optimal}\index{optimal algorithm} algorithm is one which is guaranteed to find a global minimum to this cost function; it will find an isomorphism if it exists, but still handles the problem of graph variability. As a result, however, optimal algorithms for inexact matching are significantly more expensive than their exact counterparts \cite{Conte_2004}.
Most inexact matching algorithms are therefore \emph{approximate}\index{approximate algorithm} or \emph{suboptimal}\index{suboptimal algorithm} algorithms. They only guarantee to find a local minimum of the cost function (either by optimizing it directly or by approximating it by some other cost function). This local minimum may or may not be close to the true minimum, but the cost savings of using these algorithms can outweigh this downside~\cite{Conte_2004}.
\subsubsection{Mapping-seeking algorithms} 
Finally, we distinguish between \emph{mapping seeking} algorithms, i.e., those algorithms that return a mapping between two networks, and \emph{non-mapping-seeking} algorithms that merely return a score rating the match. All exact formulations seek a mapping, and many inexact formulations do as well. Mapping-seeking inexact matching is more commonly referred to as \emph{alignment}\index{alignment} and is one of two overwhelmingly dominant comparison strategies in biological applications. Alignment is discussed in more detail in Section \ref{chapter:systems_biology}.

\subsection{Graph matching methods}

The field of graph matching is large and well-established, and we cannot hope to give a full overview of all existing techniques. For a more comprehensive investigation, we refer the reader to the definitive source on graph matching developments through 2004 \cite{Conte_2004}, a similar survey covering the subsequent ten years \cite{foggia2014graph}, and a 2018 large-scale performance comparison of graph matching algorithms \cite{carletti2018comparing} that may also be of interest. Here, we partition the field into three categories:


\begin{enumerate}
\item Exact matching methods.
\item Edit distance-based methods for optimal inexact matching. 
\item Cost function-based optimization methods for inexact matching. \end{enumerate}

We present optimization methods in Sections~\ref{sec:gedist} and~\ref{oldchapter4} and network alignment and comparison methods in Section \ref{chapter:systems_biology}. In the next two sections, we introduce the concepts of search space pruning (the most dominant approach for exact matching), a graph edit path, and its corresponding graph edit distance. Our presentation of edit distances is primarily inspired by \cite{Livi_2012} and \cite{Riesen_2009}.

\subsection{Exact matching}\index{search space pruning}

Most algorithms for exact graph matching are based on some form of tree search\index{tree search} with backtracking \cite{Conte_2004}. The process is analogous to solving a grid-based logic puzzle: we represent all possible matching options in a grid format and then rule out infeasible possibilities based on clues or heuristics about the problem. When we get to the point where our clues can no longer rule out any further possibilities, we arbitrarily choose one of the remaining possible options for a certain item. This rules out other possibilities, allowing us to use our clues again. We continue this process until we either complete the puzzle or reach a state where there are no possible solutions remaining. In the latter case, we backtrack, rule out our initial arbitrary choice, and try its alternatives until we either find a solution or exhaust all possible choices. The backtracking process is a depth-first search of the tree of possible mappings; our clues allow us to skip over branches without searching them.

The seminal algorithm for exact matching is due to Ullman~\cite{Ullmann_1976} and is applicable to both graph and subgraph isomorphism. We assume two graphs $g_1$ and $g_2$ with node counts $m$ and $n$, respectively, and assume without loss of generality that $m\leq n$. 
%
%
The Ullmann algorithm\index{Ullmann subgraph isomorphism algorithm}\index{subgraph isomorphism} uses two principles to rule out matching possibilities:

\begin{enumerate}
\item We cannot map a node in $g_1$ to one in $g_2$ that has fewer neighbors. This strategy is used to rule out matching possibilities initially and usually drastically decreases the number of possible matches at a very low computational cost.
\item We cannot map a node $v_1\in g_1$ to a node $v_2\in g_2$ unless all its neighbors have feasible matching possibilities among the neighbors of $v_2$. Testing this is called the \emph{refinement}\index{refinement} procedure, and it forms the heart of the algorithm. 
\end{enumerate}

A visual demonstration of the Ullmann algorithm can be found in the full version of this text, available at \url{https://github.com/marissa-graham/network-similarity}.

\subsection{Graph edit distance}\label{sec:gedist}


One way to measure the distance between two objects is to measure how much work it takes to turn the first into the second, and take the length of the \emph{edit path}\index{edit path}. The \emph{edit distance} between two objects is defined to be the {minimum} over the lengths of all possible edit paths between them.
%
%
For graphs, the relevant edit operations are node substitution\index{graph edit operations}, node insertion, node deletion, edge insertion, and edge deletion. Instead of simply taking the length of the edit path, however, we associate each of these operations with some nonnegative cost function $c(u,v)\in \mathbb{R}^+$ (the ``penalty" mentioned in Section \ref{section:defining_graph_matching}) which avoids rewarding unnecessary edit operations by satisfying the inequality $c(u,w)\leq c(u,v)+c(v,w)$, where $u, v$, and $w$ are nodes or edges, or sometimes null nodes/edges in the case of insertion and deletion. We also assume that the cost of deleting a node with edges is equivalent to that of first deleting each of its edges and then deleting the resultant neighborless node. 

The edit distance is then the total cost of all operations involved in an edit path, and it critically depends on the costs of the underlying edit operations \cite{Bunke_1998}. This is helpful in some cases, as it allows us to easily tweak parameters in our notion of similarity, but in others we prefer to avoid this dependence on the cost function. This is one motivation for the formulation of inexact graph matching as a continuous optimization problem, which we discuss in the next section.

Finally, we note that it was shown by Bunke that the graph isomorphism, subgraph isomorphism, and maximum common subgraph problems are all special cases of the problem of calculating the graph edit distance under certain cost functions~\cite{Bunke_1999}.

\subsection{Cost function-based methods for inexact matching}\label{oldchapter4}
As observed previously, optimal methods for inexact graph matching are expensive and typically only suitable for graphs of small size. To address this issue, Riesen and Bunke introduced an algorithm for approximating the graph edit distance in a substantially faster way~\cite{Riesen_2009}, which Serratosa published an improved variant~\cite{Serratosa_2014}. It is not the only suboptimal inexact matching method with the goal of suboptimally calculating a graph edit distance, but it provides an interesting connection between the seemingly disparate strategies of search space pruning and optimizing a cost function.
\subsubsection{The assignment problem}

The key to this connection is the idea of the assignment problem. Instead of searching the space of possible edit paths to find the graph edit distance, we approximate the graph edit distance with a solution to a certain matrix optimization problem. The following definition is due to Riesen and Bunke \cite{Riesen_2009}:

\textit{Consider two sets $A$ and $B$, each of cardinality $n$, together with an $n\times n$ cost matrix $C$ of real numbers where the matrix elements $c_{i,j}$ correspond to the cost of assigning the $i$-th element of $A$ to the $j$-th element of $B$. The \emph{assignment problem}\index{assignment problem} is that of finding a permutation $p=\{p_1,\dots,p_n\}$ of the integers $\{1,2,\dots,n\}$ which minimizes the sum $\sum_{i=1}^n c_{i,p_i}$ of the individual assignment costs.}

A brute force algorithm for the assignment problem would require a $O(n!)$ time complexity, which is impractical. Instead, we typically use the {Hungarian method}\index{Hungarian algorithm}\index{Munkres' algorithm}. This algorithm is originally due to Kuhn~\cite{Kuhn_1955} and solves the problem in maximum time $O(n^3)$ by transforming the original cost matrix into an equivalent matrix with $n$ independent zero elements which correspond to the optimal assignment pairs. The version of the algorithm described in \cite{Riesen_2009} is a refinement of the original Hungarian algorithm published by Munkres~\cite{munkres1957algorithms}.
\subsubsection{The bipartite graph matching problem}

\begin{figure}[!t]
\centering
\begin{tabular}{m{0.3\textwidth}m{0.05\textwidth}m{0.3\textwidth}}
$C = $\bordermatrix{ & 1 & 2 & 3 \cr
a & 3 & 2 & 1 \cr
b & 1 & 3 & 4 \cr
c & 2 & 5 & 2 }
 & $\Leftrightarrow$ &
\includegraphics[width=0.3\textwidth]{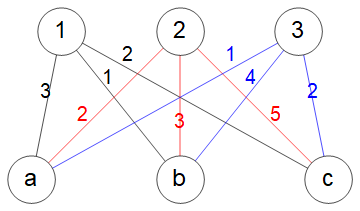} \\
\multicolumn{3}{c}{$A = \{a,b,c\},  B=\{1,2,3\}$}
\end{tabular}
\caption{Reformulating the assignment problem as that of finding an optimal matching in a bipartite graph. The edges and their weight labels in the bipartite graph are colored to make it easier to see which weights belong to which edges.}
\label{fig:bipartite_reformulation}
\end{figure}

As noted in Section \ref{section:defining_graph_matching}, we sometimes must address the question of finding a matching \emph{in} a graph\index{matching in a graph}, i.e., a set of edges without common nodes. It is straightforward to reformulate the assignment problem as one of finding an optimal matching within a {bipartite graph}\index{bipartite graph}, that is, a graph whose nodes can be divided into two disjoint independent sets such that no edges run between nodes of the same type. If $A$ and $B$ are two sets of cardinality $n$ as in the assignment problem, the elements of $A$ form one node group, the elements of $B$ form the other, and we define the edge weight between the $i$-th element of $A$ and the $j$-th element of $B$ to be the cost of that assignment, as shown in Figure~\ref{fig:bipartite_reformulation}. The assignment problem is therefore also referred to as the bipartite graph matching problem.

\subsubsection{Graph edit distance using the assignment problem}

To connect the assignment problem to graph edit distance computation, we define a cost matrix $C$ such that each $c_{i,j}$ entry represents the cost of assigning the $i$-th node in our source graph to the $j$-th node in our target graph \cite{Riesen_2009}. We can generalize this approach further to handle graphs with different numbers of nodes by using a modified version of the Hungarian method which applies to rectangular matrices \cite{bourgeois1971extension} by considering node insertions and deletions as well as substitutions. 
In this modified version of the Hungarian method, the resulting cost matrix (again, definition due to Riesen and Bunke \cite{Riesen_2009}) becomes 
\[
C = \left[
\begin{array}{cccc|cccc}
c_{1,1} & c_{1,2} & \dots & c_{1,m}     &     c_{1,-} & \infty & \dots & \infty \\
c_{2,1} & c_{2,2} & \dots & c_{2,m}     &     \infty & c_{2,-} & \ddots & \vdots \\
\vdots & \vdots & \ddots & \vdots          &     \vdots & \ddots & \ddots & \infty \\ 
c_{n,1} & c_{n,2} & \dots & c_{n,m}     &     \infty & \dots & \infty & c_{n,-} \\ \hline

c_{-,1} & \infty & \dots & \infty             &     0 & 0 & \dots & 0 \\ 
\infty & c_{-,2} & \ddots & \vdots          &     0 & 0 & \ddots & \vdots \\ 
\vdots & \ddots & \ddots & \infty           &     \vdots & \ddots & \ddots & 0\\ 
\infty & \dots & \infty & c_{-,m}            &    0 & \dots & 0 & 0 \\ 
\end{array}
\right],
\]
where $n$ is the number of nodes in the source graph, $m$ is the number of nodes in the target, and a $-$ is used to represent null values. The upper left corner of this matrix represents the cost of node substitutions, and the bottom left and top right corners represent the costs of node insertions and deletions. Since each node can be inserted or deleted at most once, the off-diagonal elements of these are set to infinity. Finally, since substitutions of null values should not impose any costs, the bottom right corner of $C$ is set to zero.

This is only a rough approximation of the true edit distance, as it does not consider any information about the costs of edge transformations. We can improve the approximation by adding the minimum sum of edge edit operation costs implied by a node substitution to that substitution's entry in the cost matrix, but we still only have a suboptimal solution for the graph edit distance problem even though the assignment problem can be solved optimally in a reasonable amount of time.

\subsubsection{Other suboptimal graph matching methods using the assignment problem}

Approximating the graph edit distance is far from the only graph matching strategy which is based around the assignment problem. Instead of a cost matrix based around the cost function of a graph edit distance measure, we can incorporate other measures of similarity or affinity between nodes. The advantage of this approach is that we can incorporate both topological and external notions of similarity into our definition. To be effective, this strategy requires a relevant source of external information, and as a result is much more prevalent in biological applications. In either case, we attempt to maximize the edges in the induced common subgraph only indirectly; a good cost function will be an effective proxy for how much a certain node pairing will contribute to a good overall mapping, but it does not directly maximize the edge preservation.
\subsubsection{Weighted graph matching vs. the assignment problem}

Most of the suboptimal graph matching methods we observed are based around either the assignment problem or around some formulation of the weighted graph matching problem. The {weighted graph matching problem}\index{weighted graph matching problem} is typically defined as finding an optimum permutation matrix which minimizes a distance measure between two weighted graphs; generally, if $A_G$ and $A_H$ are the adjacency matrices of these, both $n\times n$, we seek to minimize $||A_G - PA_HP^T||$ with respect to some norm \cite{Umeyama_1988, Koutra_2013, Almohamad_1993} or to minimize some similarly formulated energy function \cite{Gold_1996}. The specific definition depends on the technique being used to solve the optimization problem.
Weighted graph matching is an inexact graph matching method, and its techniques are generally suboptimal, searching for a local minimum of the corresponding continuous optimization problem. A wide variety of techniques are used, including linear programming \cite{Almohamad_1993}, eigendecomposition \cite{Umeyama_1988}, gradient descent \cite{Koutra_2013}, and graduated assignment \cite{Gold_1996}. Other techniques mentioned in \cite{Almohamad_1993, Umeyama_1988, Gold_1996, Conte_2004} include Lagrangian relaxation, symmetric polynomial transformation, replicator equations, spectral methods other than eigendecomposition, neural networks, and genetic algorithms.

The weighted graph matching problem is similar to the assignment problem in that we seek a permutation between the $n$ nodes of two graphs, but unlike the assignment problem, there is no need to define a cost or similarity matrix ahead of time. Instead, we directly measure the quality of a permutation assignment with respect to the structure of a graph and optimize this quantity to find our matching. This allows us to avoid relying on the heuristics inherent in any formulation of a cost or similarity matrix, but it also means we cannot easily incorporate external information into our solution of the problem. We also cannot choose to favor alignments with desirable properties other than conserved edges; for example, we may prefer a connected alignment to a more scattered one, even if it conserves fewer edges. Whether weighted graph matching techniques are preferable to assignment problem-based strategies is therefore dependent on the specific problem to be solved.

\subsection{Graph kernels and embeddings}

According to \cite{Livi_2012}, there are three main approaches to the inexact graph matching problem: edit distances, graph kernels, and graph embeddings. Edit distance-based methods are the dominant approach, but we briefly discuss the other two for completeness.

Graph \emph{embeddings}\index{graph embedding} map a graph into some high-dimensional feature space, allowing us to perform comparisons there \cite{Emmert_Streib_2011}. For example, we could identify a graph with a vector in $\mathbb{R}^n$ containing the statistics reported in Table \ref{tab:network_table} or the eigenvalues of its adjacency matrix and then compare them using the Euclidean distance. Mapping a graph into Euclidean space certainly makes comparison easier, but it is not obvious how to create a mapping that preserves graph properties in a meaningful way. Graph statistics and embeddings therefore must be experimentally shown to be useful; they may allow us to distinguish between different classes of graphs, correlate with some other desirable property, narrow down matching candidates in a large database, and so on.

Graph \emph{kernels}\index{graph kernel} are a special kind of graph embeddings in which we have a continuous map $k:\mathcal{G}\times \mathcal{G}\rightarrow \mathbb{R}$, where $\mathcal{G}$ is the space of all possible graphs, such that $k$ is symmetric and positive definite or semidefinite \cite{Livi_2012}. Creating a graph kernel would allow us to take advantage of the techniques and theory of general kernel methods, but computing a strictly positive definite graph kernel is at least as hard as solving the graph isomorphism problem \cite{Gartner_2003}. We suspect that the cost of creating a (not necessarily strictly positive definite) graph kernel with enough desirable properties to take advantage of kernel methods is prohibitive enough to make this an impractical strategy in most cases.

We note that the strategy of using the graphlet degree distribution\index{graphlet degree distribution}, which we discuss in Section \ref{chapter:systems_biology}, is a form of embedding. Furthermore, the graph kernel strategies described in the references of  \cite{Livi_2012} tend to use the assignment problem-style approach of calculating some sort of similarity notion between pairs of nodes in two graphs and then using that matrix to create the desired alignment or kernel. We therefore consider the strategies of graph kernels and graph embeddings to be part of the families of other categories which we describe in this work rather than mainstream approaches in their own right.

%
%
%
%
%
%

\section{Network Similarity in Biology}\label{chapter:systems_biology}

A fundamental goal in biology is to understand complex systems in terms of their interacting components. Network representations of these systems are particularly helpful at the cellular and molecular level, at which we have large-scale, experimentally determined data about the interactions between proteins, genes, and metabolites. In the past twenty years, there has been an explosion of availability of interaction data between biomolecules, paralleling the surge of DNA sequence information availability that was facilitated by the Human Genome Project \cite{humanGenomeProject}. Sequence information comparison tools have been revolutionary in advancing our understanding of basic biological processes, including our models of evolutionary processes and disease \cite{HGPimpact}, and the comparative analysis of biological networks presents a similarly powerful method for organizing large-scale interaction data into models of cellular signaling and regulatory machinery \cite{Sharan_2005}. %

In particular, network comparison techniques can address fundamental biological questions such as ``Which proteins, protein interactions, and groups of interactions are likely to have equivalent functions across species?", ``Can we predict new functional information about proteins and interactions that are poorly characterized, based on similarities in their interaction networks?", and ``What do these relationships tell us about the evolution of proteins, networks, and whole species?" \cite{Sharan_2006}. Comparison strategies and metrics are also key to developing mathematical models of biological networks which represent their structure in a meaningful way--a key step towards understanding them. For example, good comparison techniques allow us to model dynamical systems on biological networks \cite{Watts_1998} (e.g., the spread of infectious diseases) and to create appropriate null hypotheses for drawing conclusions about experimental networks.

We observed two overwhelmingly dominant network comparison strategies in biology applications. We  use the term {comparison} strategies and not {matching} or {alignment} strategies because--unlike pattern recognition--not all biological applications seek a mapping between two networks. This is a result of the difference between the typical networks with which each field is concerned. In pattern recognition, graphs are primarily a convenient data structure to represent the objects that we would like to compare and do not necessarily represent inherent real-world relationships. We typically have a small number of vertices and can expect a close structural match.  

On the other hand, biology considers networks which are very large, typically incompletely explored, and non-deterministic\footnote{Since  biological  networks  are  constructed  entirely  based  on  experimental  data,  edges  can  only  ever represent probabilities, and their analysis must take this into consideration.}, and for which we cannot expect a close structural match. Furthermore, we have significant external information about their vertices, such as BLAST scores\index{BLAST scores} or known functions of individual proteins, and every edge and subgraph corresponds to real-world interactions that have real-world meaning. Mapping-seeking comparison strategies for biology do not seek to give an overall measure of the similarity between two graphs, but to find regions which are conserved between two or more networks, under what Flannick et al.\ \cite{flannick2006graemlin} calls ``perhaps the most important premise of modern biology": the assumption that evolutionary conservation implies functional significance.

When comparing and analyzing biological networks as a whole, we can find meaning using strategies other than seeking a mapping between networks. These are generally based on investigating the frequencies and/or distributions of relevant subgraphs of a large network, i.e., graphlets and motifs.

\begin{figure}[t]
\centering
\includegraphics[width=\textwidth]{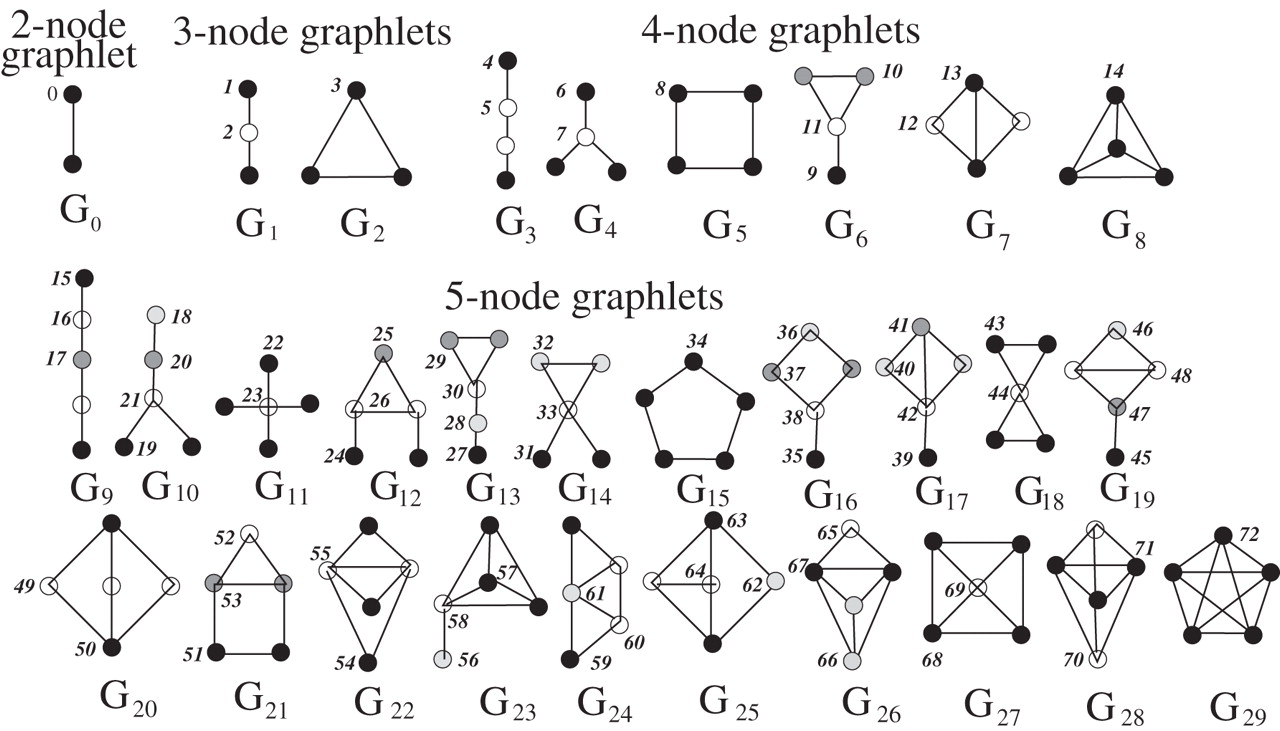}
\caption{The 73 automorphism orbits for the 30 possible graphlets with 2-5 nodes. In each graphlet, vertices belonging to the same automorphism orbit are the same shade. Figure reproduced from \cite{Przulj_2007}.}
\label{fig:graphlets}
\end{figure}

\begin{table}[H]
\centering
{\setlength\extrarowheight{1pt}\setlength{\tabcolsep}{3pt}
\begin{tabular}{|L{5cm}|c|c|c|C{2cm}|C{2cm}|}
\hline
 & Vertices & Edges & Clustering & Maximum $G_0$ degree & Maximum $G_2$ deg. \\ \hline
\textit{Mycoplasma genitalium} & 444 & 1860 &  0.758 & 66 & 1376 \\ 
Random (match degree seq.) & 444 & 1860 & 0.420 & 66 & 774  \\ 
Random (match size) & 444 & 1860 &  0.022 & 17 & 6 \\ \hline
\textit{Schizosaccharomyces pombe} & 5100 & 30118 & 0.757 & 213 & 14592 \\ 
Random (match degree seq.) & 5100 & 30118 &  0.150 & 213 & 3606 \\ 
Random (match size) & 5100 & 30118 & 0.002 & 26 & 3 \\ \hline
\end{tabular}
}
\caption{Statistics for PPI networks of two small organisms and two comparable random graphs for each. The ``clustering" value is the global clustering coefficient\index{global clustering coefficient} \cite{newman2010}, which measures the fraction of connected triplets in the network which are closed.}
\label{tab:ppi_networks}
\end{table}
\vspace{-.25cm}

\subsection{Graphlets}

\emph{Graphlets}\index{graphlets} are small connected non-isomorphic induced subgraphs of a simple undirected network \cite{Przulj_2007}, introduced to design a new measure of local structural similarity between two networks based on their relative frequency distributions.

Recall that the degree distribution measures, for each value of $k$, the number of vertices of degree $k$. In other words, for each value of $k$, it gives the number of vertices touching $k$ edges. A single edge is the only graphlet with two nodes, and we call it $G_0$. The degree distribution can therefore be thought of as measuring how many vertices touch one $G_0$ graphlet, how many vertices touch two $G_0$ graphlets, and so on.

We can generalize this idea to larger graphlets and count how many vertices touch a certain number of each graphlet $G_0, \dots, G_{29}$, where $G_0,\dots, G_{29}$ are defined as in Figure \ref{fig:graphlets}. For example, the $G_2$ distribution measures how many vertices touch one triangle, two triangles, and so on. For most graphlets larger than two vertices, {which} vertex of the graphlet we touch is topologically relevant; for example, touching the middle node of $G_1$ is different from touching an end node. This is because the end and middle vertices are in different {automorphism orbits}\index{automorphism orbits}, i.e., we can tell them apart without labeling the graph using features like their degree or the neighbors of their neighbors.

The \emph{graphlet degree}\index{graphlet degree} therefore measures how many of a certain graphlet {in a specific automorphism orbit} each vertex touches. There are 73 different automorphism orbits for the thirty graphlets with 2-5 nodes, and we therefore obtain 73 {graphlet degree distributions (GDDs)}\index{graphlet degree distribution} analogous to (and including) the degree distribution. Since these are based on the neighborhoods of each vertex, they measure the local structure of a graph.

\begin{figure}[H]
\centering
\includegraphics[width=0.9\textwidth]{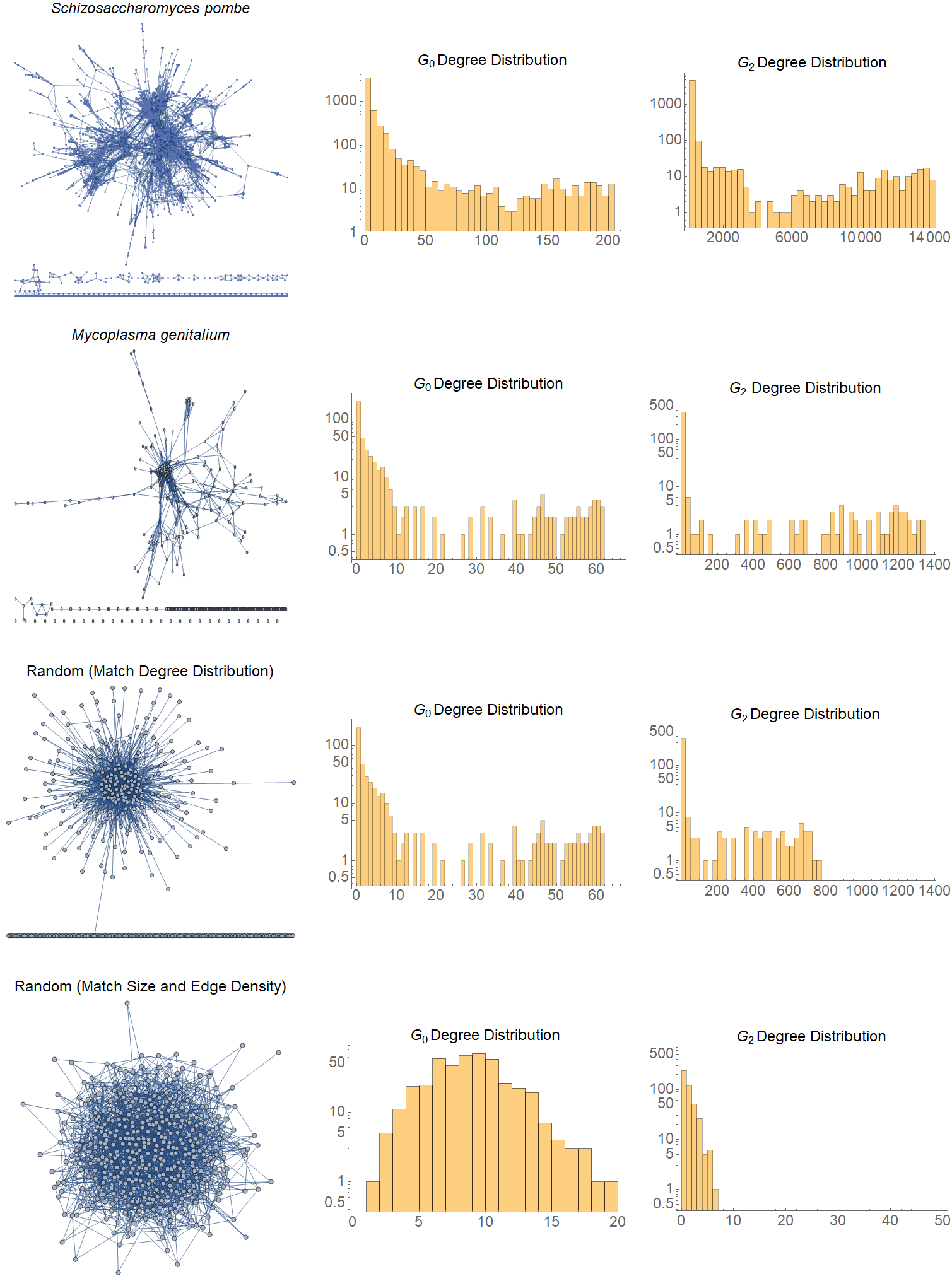}
\caption{Visualizations and $G_0$ and $G_2$ distributions for the PPI networks of \textit{S. pombe}, \textit{Mycoplasma genitalium}, and two randomly generated networks comparable to the latter.}
\label{fig:GDD_demo}
\end{figure}

An example of the $G_0$ and $G_2$ distributions for the protein-protein interaction (PPI) networks of \textit{Mycoplasma genitalium} (an STD) and \textit{Schizosaccharomyces pombe} (yeast) is shown in Figure \ref{fig:GDD_demo}, with corresponding statistics in Table \ref{tab:ppi_networks}. We obtained both networks from the STRING database \cite{szklarczyk2014string} and included links with an interaction confidence score above 950, i.e., those representing 95\% probability of an interaction between two proteins. 

When we look at the maximum $G_2$ degree for the networks in Table  \ref{tab:ppi_networks}, we can easily distinguish between a real network and a random network with the same degree sequence; we can also see different patterns in the shape of the $G_2$ distributions for the PPI networks of our two model organisms. We can guess that the varying shapes of all 73 GDDs would help us distinguish between different graphs in a meaningful way. In order to use these distributions for computational network comparison, however, we must somehow reduce this large quantity of information to a single measure. One possibility ~\cite{Przulj_2007} is to consider the Euclidean distance between each GDD for two networks, after appropriate scaling and normalization, and then take the arithmetic or geometric mean over this distance for all 73 graphlet automorphism orbits. Other methods are possible, and their suitability depends on the application in question.

\subsection{Motifs}

Network \emph{motifs}\index{motif} are similar to graphlets in that they are both small induced subgraphs of large networks. Unlike graphlets, however, the definition of motifs requires these subgraphs to be {statistically overrepresented} in the network \cite{Milo_2002}. In a random graph with the same degree sequence(s) as a real network, we are not likely to see connected triangles, for example; but as we see in Table \ref{tab:ppi_networks} and Figure \ref{fig:GDD_demo}, connected triangles appear frequently in real biological networks, associated with feedback or feed-forward loops in transcription and neural networks, clusters in protein interaction networks, and so on \cite{Berg_2004}. We can generalize further to seek \emph{topological motifs}, which are statistically overrepresented ``similar" network sub-patterns. 

Such patterns can be used as a first step towards understanding the basic structural elements particular to certain classes of networks \cite{Milo_2002}; different types of networks contain different types of elementary structures which reflect the underlying processes that generated them, and as discussed in \cite{Berg_2004}, motifs are indicative of biological functions. These methods are a useful way to distinguish patterns of biological function in the topology of molecular interaction networks from random background, but they are not well-suited for full-scale comparison of multiple networks \cite{Przulj_2007}. They are sensitive to the choice of random network model used to determine statistical significance, and they ignore subgraphs with low or average frequency. These low-frequency subgraphs may still be functionally important, especially if such subgraphs are consistently seen across multiple real networks despite occurring rarely within any individual one. Graphlets are one way to address this issue; alignment strategies are another. We note that \cite{Przulj_2007} defines graphlets for undirected graphs only, while the motifs discussed in \cite{Milo_2002} and \cite{Berg_2004} are primarily directed. In both cases, however, the definition does not include multiedges or self-loops.


Subgraph isomorphism is a computationally expensive problem, which makes exhaustively finding all occurrences of small isomorphic or near-isomorphic subgraphs in a network infeasible for all but small biological interaction networks. As a result, motif search in large networks (which requires an exhaustive search over an entire ensemble of random graphs in order to determine statistical significance) is generally limited to subgraphs of at most five nodes. Graphlet statistics are similarly expensive to compute. In order to process the interaction networks of higher organisms, search heuristics and estimation procedures must be used.

\subsubsection{Netdis}

Netdis \cite{Ali_2014} is an alternative method for network comparison which is based on counting the occurrences of small subgraphs in a larger network. For each vertex, the number of occurrences of each possible graphlet of 3-5 nodes is counted in a neighborhood of radius two around it. Each vertex is thereby associated with a vector of graphlet counts, which is normalized with respect to a proxy for the counts we would expect to see in a suitable random model. 

These centered counts are then combined into an overall statistic, which is used in \cite{Ali_2014} to correctly separate random graph model types and to build the correct phylogenetic tree of species based on their protein interaction networks, showing that Netdis is a relevant comparison method for large networks. The method is also highly tractable; since subgraphs are only searched for in a given neighborhood, computational complexity grows about linearly with the number of vertices in a network if neighborhood sizes stay relatively small, and Netdis is therefore well-suited for full-scale comparison of many large networks.

While Netdis uses graphlet counts, it is not a generalization or special case of graphlet degrees. When counting graphlets in the two-step neighborhood of a vertex, we only need to consider the 29 possible graphlets of 3-5 nodes, rather than their 72 possible automorphism orbits. The graphlets a vertex touches will not always be present in its two-step neighborhood, and there are often graphlets in the two-step neighborhood of a vertex which do not directly touch the vertex itself.

\subsection{Local and global network alignment}

In graph matching and in subgraph counting, the mappings found by an algorithm are not the primary result of interest. When we calculate graphlet statistics and network motifs, while graph matching is performed, the subgraphs involved are so small that the computational difficulty is in enumerating the relevant subgraphs, not in matching them to each other. Inexact graph matching methods are used on graphs with enough nodes to make an exhaustive search difficult, so the matching method we use matters, but the mapping between two networks does not typically give us relevant real-world insights about the structures they represent.

When comparing large biological networks, on the other hand, mappings give us real-world insights, and we frequently seek regions of similarity and dissimilarity in networks with thousands of vertices and tens of thousands of edges. Just as longer DNA sequences which are conserved across species indicate functional significance and can help classify evolutionary relationships, larger subnetworks of biomolecular interactions which are conserved across species are likely to represent true functional modules\cite{Sharan_2006} and give insight into evolutionary processes \cite{Ali_2014}. To find these conserved regions, we seek an \emph{alignment}\index{alignment}, or a mapping between vertices in two or more networks which approximates the true structure they have in common. Alignment algorithms may not find an isomorphism if one exists, but this is not important when the networks involved are never isomorphic.

Alignment strategies for biological applications fall into two categories: local alignment, and global alignment. In both cases, we seek to find regions of similarity between networks, and mappings do not need to be defined for every vertex in each network. 
%
\begin{figure}[t]
\centering
$\begin{array}{ccc}
\includegraphics[width=0.4\textwidth]{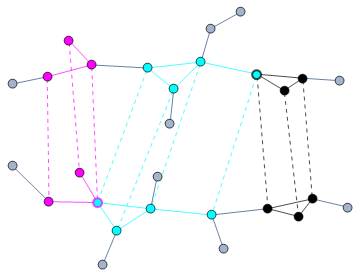}
&\quad\quad&\includegraphics[width=0.4\textwidth]{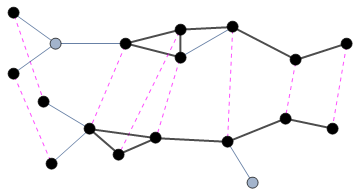}
\end{array}$
\caption{Local alignment of a network (left panel) and global alignment of a network (right panel). In local alignment, vertices may be used for multiple ``pieces" of the overall mapping, i.e., the mapping is not required to be one-to-one. In global alignment, not all vertices must be mapped to a vertex in the other network, but the mapping must be one-to-one in both directions when they are.}
\label{fig:global_alignment}
\end{figure}

In \emph{local alignment}\index{local alignment}, we seek local regions of isomorphism between the two networks, where each region implies a mapping which is independent of the others. These mappings do not need to be mutually consistent; that is, we can map a vertex differently under different regions of the mapping, as illustrated in Figure \ref{fig:global_alignment}. We can choose a locally optimal mapping for each region of similarity, even if this results in overlap. In \emph{global alignment}\index{global alignment}, by contrast, we define a single mapping across all parts of the input, even if it is locally suboptimal in some regions of the networks \cite{Singh_2007}. 

To understand the distinctions between these two alignment strategies, we must understand the motivations behind their development, and the biological implications of their results~\cite{flannick2006graemlin}. The study of biological \emph{networks} has been guided first and foremost by the study of biological \emph{sequences}, and the development of biological network alignment closely parallels that of sequence alignment\index{sequence alignment}. Sequence alignment is based on the assumptions that:
\begin{enumerate}
\item\label{first} Patterns which occur frequently (motifs) are likely to have functional significance. \item\label{second} Sequence regions which are conserved across multiple species are likely to have
 functional significance.
\item \label{third}The degree to which sequences differ is related to their evolutionary distance.
\end{enumerate}

The study of~\eqref{first} is roughly the domain of local sequence alignment, while the study of~\eqref{second} and~\eqref{third} is roughly the domain of global sequence alignment.  Local network alignment, by analogy, searches for highly similar network regions that likely represent conserved functional structures (i.e., evolutionarily conserved building blocks of cellular machinery), which often results in relatively small mapped subnetworks and in some network regions not being a part of the alignment. Global network alignment, on the other hand, looks for the best superimposition of the entire input networks  (i.e., a single comprehensive mapping between the sets of protein interactions from different species), which typically results in large but suboptimally conserved mapped subnetworks \cite{guzzi2017survey}.

Sequence alignment cannot be perfectly generalized to network alignment; sequence complexity is linear, while network complexity is combinatorial. An alignment of two sequences of length $m$ and $n$ can be done in $O(mn)$ time using the Needleman-Wunsch algorithm \cite{Wunsch_time_is_over}\index{Needleman-Wunsch algorithm}, but the maximum common induced subgraph problem--the closest analogue of sequence alignment for networks--is known to be NP-hard. As a result, the analogy to sequence alignment must be compromised for the sake of tractability. This is least necessary in the case of very small subgraphs. Searching exhaustively for very small subgraphs in a network is computationally expensive, but tractable. As a result, network motifs are a fairly straightforward generalization of the idea of sequence motifs.

For slightly larger subgraphs, a common compromise is to search for a predetermined pattern, rather than for {all} frequently occurring patterns. The number of possible patterns on 2-5 vertices is small enough to make searching for all of them feasible (i.e., graphlets and motifs), but one can also search a less well-studied network or the network of another species for a certain pathway or module known to be significant. Another compromise is to construct an alignment based on a deterministic notion of which nodes are similar and which edges are conserved. Since we have significant amounts of external information about the vertices, this can yield useful results. Not all local alignment algorithms in our references make the compromise of searching only for predetermined patterns--although this is a common strategy--but they \emph{do} all define alignments deterministically through external information. Local alignment defines a less constrained problem than global alignment, one too broad for network topology to answer on its own.

Global alignment, on the other hand, makes the compromise of investigating the assignment problem rather than the maximum common induced subgraph problem. The assignment problem has the advantage of being fully solvable in low polynomial time via the Hungarian algorithm, which allows the handling of much larger networks. The tradeoff is that the conserved networks we find are highly dependent on our cost function for mapping vertices onto each other, and we must use heuristic measures to determine the quality of the resulting assignment. The cost function is generally based on a combination of external (i.e., sequential and/or functional) and topological information about the vertices in a network. Unlike local alignment, the use of external information is frequently optional, and it is never the primary driver of the results.

At the time IsoRank\index{IsoRank} \cite{Singh_2007} was introduced in 2007, the global network alignment problem had received little attention in the literature. After IsoRank was introduced, however, global alignment began to receive significant attention, and most subsequent papers on biological network alignment address global alignment. The transition from local to global alignment seems to have occurred as biological network comparison strategies increasingly sought to learn from topological network structure rather than deterministically attempting to generalize sequence alignment, and as increasingly accurate biological network data and increased computational resources for network analysis became available.

\subsection{Local alignment algorithms}

We now present a few of the most important algorithms for local network alignment.
\subsubsection{PathBLAST and NetworkBLAST} The PathBLAST algorithm\index{PathBLAST}~\cite{kelley2004pathblast, Sharan_2005, Sharan_2006} and its successor NetworkBLAST~\cite{kalaev2008networkblast} both search for query pathways or query networks in a larger target network (or in the case of NetworkBLAST, multiple target networks). PathBLAST is primarily designed to search for conserved pathways, and it handles query networks by searching for conservation of the pathways in the query network. NetworkBLAST\index{NetworkBLAST} extends the algorithm to multiple networks by searching in an alignment network of multiple species, which it constructs deterministically by matching vertices according to the sequence similarity of their corresponding proteins and defining fixed conditions under which an edge is considered to be conserved rather than searching over the space of possible alignments.

\subsubsection{Graemlin} While PathBLAST can search within multiple networks, it cannot effectively find conserved modules of arbitary topology within an arbitrary number of networks. Graemlin (General and Robust Alignment of Multiple Large Interaction Networks)~\cite{flannick2006graemlin}\index{Graemlin} addresses this problem by first defining an alignment between networks deterministically. In order to effectively search for an alignment across multiple networks, it uses a progressive strategy of successively aligning the closest pairs of networks using a phylogenetic tree\index{phylogenetic tree}. At each level of the tree, it pairwise aligns the most closely related species and uses the alignment results as the ``parent" of each pair until it reaches the root of the tree. Although this is an effective and scalable method of searching for conserved structures across multiple networks, it depends on prior knowledge of the relationships between species. It cannot infer a phylogenetic tree solely from topological data as Netdis does.

A successor to the Graemlin algorithm, Graemlin 2.0 \cite{flannick2009automatic}, was published in 2009.

\subsubsection{M$_\text{A}$WIS$_\text{H}$} With the goal of extending the concepts of matches, mismatches and gaps in sequence alignment to networks, M$_\text{A}$WIS$_\text{H}$ is an evolutionarily-inspired framework for the local alignment problem~\cite{Koyuturk_2006}. In this algorithm an alignment graph  is constructed between two networks, where each node represents a pair of ortholog\index{ortholog} proteins\footnote{Proteins in different species derived from a common ancestor gene.}, and edges between two pairs of orthologs are deterministically assigned weights which encode evolutionary information about the proteins in each pair. Instead of searching for small known patterns, however, the goal in this approach is to solve the {maximum weight induced subgraph problem}\index{maximum weight induced subgraph problem} which can be stated as:

\textit{Given a weighted graph $G(V,E)$ with edge weights $w(v,v')$ for vertices $v,v'\in V$ and a constant $\epsilon>0$, find a subset of vertices $V^*\subset V$ such that the sum of the weights of the edges in the subgraph induced by $V^*$ is at least $\epsilon$; that is, $\sum_{v,v'\in V^*} w(v,v') \geq \epsilon$.}

We note that for this to be a nontrivial problem, positive and negative edge weights in $G$ must exist; else the obvious solution is to simply choose all nodes in $G$. In the case of the alignment graph constructed in~\cite{Koyuturk_2006}, these negative edge weights are the result of evolutionarily-inspired matching penalties. M$_\text{A}$WIS$_\text{H}$ is also technically a decision problem, not an optimization problem; we aim to reach a certain goal sum of edge weights, not the maximum possible sum.

If we require that no edges in the result share a common vertex, this becomes the problem of finding a high-scoring matching within a graph, which is closely related to the bipartite graph matching problem and therefore the assignment problem. As stated, M$_\text{A}$WIS$_\text{H}$ is NP-complete, which can be shown by reduction from the maximum clique problem for the alignment graph defined by~\cite{Koyuturk_2006} (for a non-complete graph, a maximum clique will not necessarily be a solution). However, like the global alignment strategies we will discuss shortly, a reasonable solution can be found by seeding an alignment at high scoring nodes and growing it in a greedy manner.

\subsection{Global alignment algorithms}

In this section, we present alignment algorithms in the field of pattern recognition as well as biology in order to facilitate comparison of their methods and highlight the distinctions between the two fields. This is not meant to be a comprehensive overview, but we have included the most notable algorithms presented and discussed in the high centrality papers of our citation network dataset.

We previously introduced the idea of the assignment problem, in which we seek to find an overall assignment with a minimum total cost given a matrix whose entries represent the cost of assigning vertices in one network to vertices in another. In Section~\ref{oldchapter4}, this strategy was used to approximate the graph edit distance, but some formulation of the assignment problem is used in all global alignment algorithms. In order to find an alignment using the assignment problem, we must:

\begin{enumerate}
\item \label{ga1} Construct a cost matrix.
\item \label{ga2} Use that cost matrix to construct a mapping.
\end{enumerate}

The canonical example for \eqref{ga2} is the Hungarian method. Strategies for \eqref{ga1} vary significantly, but are usually based on some combination of topological and external information about the vertices in a network. We give a summary of the cost matrix and mapping construction strategies for several global alignment algorithms in Table \ref{tab:alignment_algorithms}.

\subsubsection{Non-biology methods}
Recall that an approximation of the graph edit distance can be found by searching for an optimal matching within a matrix of the costs of specific edit operations. An improved variant of Riesen and Bunke's 2009 algorithm \cite{Riesen_2009} for doing so  was introduced by Serratosa in 2014~\cite{Serratosa_2014}, which uses the same modified Hungarian method, but defines a different and smaller matrix cost in the case where edit costs result in an edit distance which is an actual distance function; that is, costs are nonnegative, substitution of identical-attribute nodes has zero cost, insertion and deletion have the same cost, and substitution costs no more than performing both an insertion and a deletion.

We saw one other assignment problem-style method from computer scientists \cite{Jouili_2009}, which uses a basic notion of node similarity and the Hungarian method to obtain reasonable but unimpressive results. More notable is the fact that it was published in 2009, two years after the introduction of IsoRank; it does not seem that the insights gained from the study of global alignment methods in biology were widely known by computer scientists at that time. This was less the case as of 2014, when Kollias et al.\ \cite{Kollias_2014} used an adapted, parallelized version of IsoRank to perform  global alignment of networks two orders of magnitude larger than previously possible (up to about a million nodes). Overall, however, the influence of biological strategies in computer science overall seems to be limited.

\begin{table}[H]
\centering
{\setlength\extrarowheight{2pt}\setlength{\tabcolsep}{3pt}\fontsize{10}{12}\selectfont
\begin{tabular}{|L{3cm}|l|L{5cm}|L{5cm}|}
\hline
\textbf{Biology} & \textbf{Year} & \textbf{Similarity Scoring} & \textbf{Alignment Construction} \\ \hline
IsoRank \cite{Singh_2007} & 2007 & Convex combination of external information and eigenvalue problem-based topological node similarities  & Maximum-weight bipartite matching OR Repeated greedy pairing of highest scores \\ \hline

Natalie \cite{Klau_2009} & 2009 & Convex combination of external info-based node mapping scores and topology-based edge mapping scores & Cast as an integer linear programming problem and use Lagrangian relaxation \\ \hline

GRAAL \cite{Kuchaiev_2010} & 2010 & Convex combination of graphlet signatures and local density & Greedy neighborhood alignment around highest-scoring pairs \\ \hline

PINALOG \cite{phan2012pinalog} & 2012 & Only sequence and functional similarity of proteins initially, but includes topological similarity for extension mapping & Detect communities, pair similar proteins from communities, extend the mapping to their neighbors  \\ \hline

GHOST \cite{Patro_2012} & 2012 & Eigenvalue distributions of appropriately normalized neighborhood Laplacians & Seed-and-extend with approximate solutions to the QAP, then local search step \\ \hline

SPINAL \cite{aladaug2013spinal} & 2013 & Convex combination of sequence similarity and neighbor matching-based topological similarity  & Seed-and-extend with local swaps \\ \hline

NETAL \cite{Neyshabur_2013} & 2013 & Update an initial scoring based on the fraction of common neighbors between matched pairs in its corresponding greedy alignment & Repeated greedy pairing of highest scores, while updating expected number of conserved interactions \\ \hline

MAGNA \cite{Saraph_2014} & 2014 & Any & Improve a population of existing alignments with crossover and a fitness function \\ \hline

Node fingerprinting \cite{radu2014node} & 2014 & Minimize degree differences and reward adjacency to already-matched pairs & Progressively add high-scoring pairings to an alignment and update scores\\ \hline\hline
\textbf{Non-biology}  & \textbf{Year} & \textbf{Similarity Scoring} & \textbf{Alignment Construction} \\ \hline

Node signatures \cite{Jouili_2009} & 2009 & Vertex degree and incident edge weights & Hungarian method \\ \hline

Graph edit distance approximation \cite{Riesen_2009} & 2009 & Edit costs (vertex insertions, substitutions, deletions) & Generalized (non-square) Munkres' algorithm \\ \hline
Modified GED approximation \cite{Serratosa_2014} & 2014 & Modification of edit costs when edit distance is a proper distance function & Generalized (non-square) Munkres' algorithm \\ \hline 
\end{tabular}
}
\vspace{0.2cm}
\caption{A broad summary of alignment algorithms discussed in this section. The distinctions between the various topological similarity scores used are discussed in each algorithm's individual section.}
\label{tab:alignment_algorithms}
\end{table}

\subsubsection{IsoRank}

IsoRank~\cite{Singh_2007} was the pioneering global network alignment method for the alignment of two networks, and was extended to the alignment of multiple networks in~\cite{Singh_2008}. It has remained a standard benchmark for the performance of subsequent algorithms. 

IsoRank calculates a similarity score between nodes by linearly interpolating between sequence similarity scores of proteins and topological similarity scores. The topological similarity score between vertex $i$ in the source network $V_1$ and vertex $j$ in the target network $V_2$ is defined to be the sum of the similarity scores for their neighbors, proportional to the total number of possible neighbor pairings. That is, we solve the eigenvalue problem \[R_{ij} = \sum_{u\in N(i)} \sum_{v\in N(j)} \frac{R_{uv}}{|N(u)||N(v)|}, \text{    } i\in V_1, j\in V_2, \] where $N(u)$ is the number of neighbors of vertex $u$. The overall similarity score between vertices $i$ and $j$ is then the solution of the eigenvalue problem \[R = \alpha AR + (1-\alpha) E, \text{    }\alpha \in [0,1], \] where $E$ is a normalized vector of pairwise sequence similarity scores and $A$ is a doubly indexed $|V_1||V_2|\times |V_1||V_2|$ matrix where $A[i,j][u,v] = 1/[N(u)N(v)]$ if there is an edge from vertex $i$ to $u$ in $V_1$ and from vertex $j$ to $v$ in $V_2$, and zero otherwise; $A[i,j][u,v]$ refers to the entry at row $(i,j)$ and column $(u,v)$. The parameter $\alpha$ controls the weight of the topological data compared to the sequence data in the overall similarity scores.

This eigenvalue problem is solved via the power method, and two methods are presented to construct an alignment~\cite{Singh_2007}: construction of the maximum-weight bipartite matching, and a greedy method which repeatedly removes the highest-scoring pairs from consideration until the alignment is finished (which sometimes performs even better than the more principled algorithm). Once all nodes are aligned, the conserved edges are simply those whose endpoints in each network are both paired to each other in the mapping. 

IsoRankN~\cite{liao2009isorankn} is a successor to IsoRank which allows for global alignment of multiple protein protein interaction networks.

\subsubsection{Natalie}

In 2009 the software package Natalie~\cite{Klau_2009} was introduced, which combined the maximum structural matching formulation for pairwise global network alignment with a Lagrangian relaxation-based algorithm for solving it. Given two networks $G_1(V_1,E_1)$ and $G_2(V_2,E_2)$, a scoring function $\sigma:V_2\times V_2\rightarrow \mathbb{R}_{\geq 0}$ for mapping individual nodes onto each other, and a scoring function $\tau:(V_1\times V_2)\times (V_1\times V_2)\rightarrow\mathbb{R}_{\geq 0}$ for mapping pairs of nodes (i.e., edges) onto each other, a \textit{maximum structural matching} of $G_1$ and $G_2$ is defined to be a mapping $M=\{M_i\}_{i=1}^n$ between the nodes (where each $M_i$, $i\in \{1,\dots,n\}$ and $n\leq \min\{|V_1|,|V_2|\}$ is a unique pair of nodes, one from each graph) which maximizes

\[s(M) = \sum_{i=1}^n \sigma(M_i) + \sum_{i=1}^n \sum_{j=i+1}^n \tau(M_i,M_j). \]

The maximization of $s(M)$ is then cast as a non-linear integer programming problem, which is reformulated as an integer linear program using Lagrangian decomposition and solved via Lagrangian relaxation. Like IsoRank, this initial paper is primarily a proof of concept. The $\sigma$ and $\tau$ functions used for the  Natalie algorithm are simplistic; $\sigma$ is defined to be $-\infty$ for proteins which are not potential orthologs according to an arbitrary sequence similarity threshold and zero otherwise, while $\tau$ is defined to be one for vertex pairs that correspond to an edge in both $V_1$ and $V_2$ and zero otherwise. An improved algorithm using the same integer linear programming framework was published in 2011 and made available as Natalie 2.0~\cite{el2011lagrangian}.
\subsubsection{GRAAL}

Introduced in 2010~\cite{Kuchaiev_2010}, GRAAL (GRAph ALigner) is unique in that it incorporates only topological information into its node similarity scores. These similarity scores are based on the \emph{graphlet degree signature} of each node, which is simply a vector of the number of each type of graphlet that the node touches. As with a graphlet degree distribution, distinctions are made between different automorphism orbits of each graphlet. Each of these 73 orbits is assigned a weight $w_i$ that accounts for dependencies between orbits\footnote{For example, differences in counts of orbit 3 will imply differences in counts of all orbits that contain a triangle, and it is therefore assigned a higher weight.}, and the \emph{signature similarity} between nodes $u\in G$ and $v\in H$ is then

\[D(u,v) = \frac{1}{\sum_{i=0}^{72} w_i} \left[ w_i \times \frac{|\log(u_i + 1) - \log(v_i+1) |}{\log (\max\{u_i, v_i\} + 2)} \right], \]
where $u_i$ is the number of times a node $u$ is touched by orbit $i$ in a graph $G$. The overall similarity between two nodes is then their signature similarity, scaled according to their relative degrees in order to favor aligning the densest parts of the networks first.

To construct an alignment using this similarity score matrix, GRAAL chooses an initial high-scoring pair and then builds neighborhoods of all possible radii around each member of the pair. These are aligned using a greedy strategy. If this does not result in a match for all the vertices in the smaller network, the same strategy is repeated for the graph $G^2$ (whose edges run between nodes connected by a path of length up to 2 in $G$), $G^3$, and so on until all the nodes in the smaller network are aligned.

This has the advantage of being well suited for networks of very different size, for which a typical greedy alignment such as that used in IsoRank is not likely to produce a connected component in the larger graph. Most global alignment strategies published after GRAAL use some variation on a seed-and-extend-neighborhoods strategy like this one in order to favor connected components in the alignment result. The authors show greatly improved results compared to IsoRank for edge conservation and connected component size in the alignment of PPI networks for yeast and fly.

Many variants within the GRAAL family have been introduced, including H-GRAAL (Hungarian-based GRAAL) in 2010 \cite{milenkovic2010optimal}, MI-GRAAL (Matching-based Integrative GRAAL) in 2011 \cite{kuchaiev2011integrative}, C-GRAAL (Common neighbor-based GRAAL) in 2012 \cite{memivsevic2012c}, and L-GRAAL (Lagrangian GRAphlet-based network ALigner) in 2015 \cite{malod2015graal}.

\subsubsection{PINALOG}

Introduced in 2012~\cite{phan2012pinalog}, PINALOG (Protein Interaction Network ALignment through Ontology of Genes) computes a global alignment between PPI networks by detecting communities within each network by merging adjacent cliques, mapping highly similar proteins from these communities onto each other using the Hungarian method, and finally extending the mapping to these highly similar proteins' neighbors to obtain the remainder of the alignment. The initial similarity definition of proteins is based solely on sequence and functional similarity, but the extension of the mapping to the neighbors of similar communities incorporates topological similarity as well to get a matrix of similarity scores, from which optimal pairings are selected using the Hungarian method.

The authors show PINALOG to conserve more interactions than IsoRank, but fewer than MI-GRAAL, and to conserve many more interactions with functional similarity than either. This is unsurprising, given PINALOG's incorporation of functional similarity scores. 

\subsubsection{GHOST}

Introduced in 2012~\cite{Patro_2012}, GHOST\footnote{According to the author, this is wordplay based on ``spectral" rather than an abbreviation or acronym, and is intended to allow colloquial usage of the name as a verb as is common with the BLAST algorithm.}  is a pairwise network alignment strategy which interpolates between topological and sequence distance information to get its overall node distance scores. Its topological distance scores are based on the density functions of the spectra for the normalized Laplacian of various-radius neighborhoods of a given vertex. We use density and not the spectra themselves because the length of each spectrum varies with the size of the neighborhood that produces it. The distances between spectral densities for two vertices are then averaged over several neighborhood radii to produce the final topological distance between them.

To align two networks, GHOST seeds regions of an alignment with close pairs of nodes from the two networks and then extends the alignment around their respective neighborhoods. Neighborhoods are matched according to an approximate solution to the quadratic assignment problem. 

This process continues until all nodes in the smaller network have been aligned to a node in the larger network. Finally, GHOST explores regions of the solution space around the initial result in hopes of a better solution.

The authors found that GHOST computes alignments of good topological and biological quality between different species. IsoRank and MI-GRAAL generally achieve only one or the other, and while Natalie 2.0 improves on both IsoRank and MI-GRAAL in this regard, GHOST dominates it in topological quality when biological quality levels are the same. The most distinct advantage of GHOST is its robustness to noise; it effectively maintains high node and edge correctness in its alignments of an increasingly noisy yeast PPI network to the original version, while the correctness of other algorithms deteriorates. 

\subsubsection{SPINAL}

In SPINAL (Scalable Protein Interaction Network ALignment), introduced in 2013 \cite{aladaug2013spinal}, similarity scores are a convex combination of a topological similarity score and sequence similarity score between a pair of proteins. Topological similarity scores are based on maximum potential conserved edges between neighbors of a potential matching pair, rather than simply being scaled by the product of their degrees as in IsoRank. The score matrix is calculated iteratively using a gradient-based method.

The alignment is constructed with a seed-and-extend method. The algorithm seeds the alignment at the highest-scoring unaligned pairs and then grows it in connected components
around these seed pairs by constructing a maximum weight matching for their neighbors (i.e., a local solution to the bipartite graph matching problem). Finally, it checks for local swaps which improve the result.

The authors extensively compare SPINAL to IsoRank and MI-GRAAL, showing improved accuracy performance on the PPI networks for various organisms and noting SPINAL's reasonable running times compared to IsoRank and MI-GRAAL.

\subsubsection{NETAL}

NETAL (NETwork ALigner) \cite{Neyshabur_2013}, published in 2013, uses a typical strategy for defining similarity scores, iteratively updating an initial scoring based on the fraction of common neighbors between matched pairs in its corresponding greedy alignment. The scoring schema is defined for both topological and biological information, although the authors do not use the biological score matrix in the calculation of their results. NETAL's alignment construction is also typical; high-scoring node pairs are iteratively added to the alignment while their corresponding rows and columns are removed from the similarity scoring matrix. 

The novel feature of NETAL is the concept of an interaction score matrix, which approximates the expected number of conserved edges obtained from aligning a certain vertex pair and which is updated as we add node pairs to the alignment. The overall similarity matrix at each stage is then a convex combination $A = \lambda T + (1-\lambda) I$ of the topological similarity matrix $T$ and the interaction score matrix $I$. 

The authors compare NETAL to IsoRank, GRAAL, and MI-GRAAL, using $\lambda=0.0001$ (about the inverse of the number of nodes in the larger network). They show improved results with respect to edge correctness and with respect to the largest common connected (not necessarily induced) subgraph in the result as well as improved robustness to noise compared to MI-GRAAL. The main advantage of NETAL is its speed. Its time complexity\footnote{Assuming $m\simeq n \log n$, which is reasonable in the case of biological networks.} is $O(n^2\log^2n)$\, compared to $O(n^5)$ for the GRAAL family. As a result, NETAL's runtime was hundreds of times lower than that of MI-GRAAL and GRAAL for the experiments performed.

\subsubsection{MAGNA}

Introduced in 2014~\cite{Saraph_2014}, MAGNA (Maximizing Accuracy in Global Network Alignment) is a technique for improving upon existing alignments which can be generated using any method. Its key observation is that existing alignment methods align similar nodes in hopes of maximizing the number of edges in a final alignment; edge conservation is not part of the alignment process.

The authors use a genetic algorithm to improve upon a population of existing alignments, where the most ``fit" alignments maximize the edge conservation in both the source and target networks being aligned. To achieve this, their fitness function penalizes both the mapping of sparse regions to dense regions and the mapping of dense regions to sparse regions, unlike previous alignment quality scoring methods. They also introduce a novel crossover method which takes the midpoint of the shortest transposition path between two parent alignments. Parent alignments are sampled from a probability distribution of the fitness of all alignments in the population.

The authors show MAGNA to improve the results of initial populations of alignments generated randomly and by IsoRank, MI-GRAAL, and GHOST. Overall, it outperformed all benchmark methods with respect to both node and edge conservation, and topological and biological alignment quality. A successor to MAGNA, MAGNA++ \cite{vijayan2015magna}, was published in 2015.

\subsubsection{Node Fingerprinting}

Node fingerprinting was introduced in 2014 \cite{radu2014node}. It aims to quickly compute accurate alignments between two networks in a parallelizable manner without relying on external information or on tunable network alignment parameters that introduce an increased computational overhead. While the approach allows for the inclusion of external information, the authors run their experiments without it to avoid the circularity of using sequence information both to carry out and to validate an alignment.

Like SPINAL and NETAL, the similarity scores used in node fingerprinting are updated throughout the process of constructing an alignment. These {pairing scores} are based on the relative differences in the in and out degrees (or simply the degree for an undirected network) of the neighbors of a potential pair to be matched; this difference is meant to be minimized. The scoring function adds a bonus for node pairings that are adjacent to already mapped node pairs and a penalty for nodes with differing in or out degrees. The algorithm then repeatedly adds the node pairings with above average scores to the alignment and recalculates pairing scores until a complete alignment has been reached.

The authors run experiments on both real and synthetic data and show equal or improved accuracy compared to IsoRank, MI-GRAAL, and GHOST, especially for large networks. In some experiments using smaller \textit{Human Herpesvirus} networks, GHOST or MI-GRAAL is able to outperform NF, but at a greatly increased runtime and memory cost. The advantage of this method is thus primarily in the analysis of very large networks, where it is able to take advantage of increased structural information at a low overall computational cost.

\subsubsection{Other algorithms}

The following algorithms may also be of interest to the reader: 
\begin{itemize}
\item HopeMap (2009) \cite{tian2009pairwise}
\item Topac (2012) \cite{guelsoy2012topac}
\item GEDEVO (2013) \cite{ibragimov2013gedevo}
\item NetCoffee (2013) \cite{hu2013netcoffee}
\item PI-SWAP (2013) \cite{chindelevitch2013optimizing}
\item SMETANA (2013) \cite{sahraeian2013smetana}
\item HubAlign (2014) \cite{hashemifar2014hubalign}
\item CytoGEDEVO (2015) \cite{malek2015cytogedevo}
\item Great (2015) \cite{crawford2015great}
\item OptNetAlign (2015) \cite{clark2015multiobjective}
\item WAVE (2015) \cite{sun2015simultaneous}
\item SANA (2017) \cite{mamano2017SANA}
\item INDEX (2017) \cite{mir2017index}
\end{itemize}

We also note that a benchmark comparison of IsoRank, Natalie 2.0, GRAAL, C-GRAAL, MI-GRAAL, PINALOG, GHOST, SPINAL, and NETAL was published by Clark and Kalita \cite{clark2014comparison} in 2014, who also collaborated with Elmsallati to publish a survey of global PPI network alignment techniques as of 2016 \cite{elmsallati2016global}.


\section{Conclusion}\label{sec:conclusion}

In this work, we set out to present an integrative overview of network comparison and matching in pattern recognition (computer science) and in systems biology, and we conclude here by discussing the high-level distinctions between the two fields and potential cross-applications for further study. 

In pattern recognition, we followed methods from the strictest possible definition of similarity between two graphs, which we determine by way of exact matching, through edit distances to approximate formulations and the assignment problem. We then introduced the differences in problem types between pattern recognition and biology, and followed biological network comparison strategies from graphlets and motifs to local and then global alignment.

A clear theme in both fields is the extent to which the graphs and networks being investigated inform the approaches used. Calculating basic statistics and counting small patterns gives us meaningful results in the context of large real-world networks, but is unhelpful for finding a small target graph in a large database. Alignment strategies in pattern recognition do not even consider the idea of incorporating external information into their definition of node similarity, while in biological applications, it is indispensable. In pattern matching, the primary measure of quality for a mapping is how well it tells us whether two graphs are isomorphic. In biology, however, alignments seek regions of similarity to find functional significance; even if we could easily find subgraph isomorphisms between two networks, they would be useless if they failed to reflect the conserved structures in the two species those networks represent.

How can we connect the ideas in these fields, despite their disparate problem types and methods, and how might they learn from each other? 

The concept of graphlets could potentially be highly applicable to pattern recognition applications and particularly towards the task of searching for a close match in a large database. In Figure \ref{fig:GDD_demo}, we saw how different two networks can be in one graphlet degree distribution despite being completely identical in another. It would be quite difficult for two graphs, no matter how large or small, to match each other precisely in all 73 without being isomorphic. Enumerating the graphlet counts in a small graph and storing even a few relevant statistics from each distribution could provide an efficient way to identify isomorphism candidates in a large database. At the time of this writing, this question does not appear to have been addressed in the pattern recognition literature; the term ``graphlets" is used, but not in the same sense. Similar ideas have likely been explored, but the specific details and theory of and available algorithms for graphlets as we have presented them could be a useful resource.

In larger networks, graphlet degree distributions could inspire random network models, just as the degree distribution inspired the idea of ``scale-free" networks and corresponding generation models for them. When properties such as scale-free, small-world, and clustering appear in real-world networks across disparate applications \cite{jackson2005economics}, we can gain better insights into how those properties arise, improve our random network models, and gain a better understanding of what distinguishes different categories of networks. 

Analysis strategies for biological networks could also be useful for the analysis of social networks; the two types have similar properties, and they are therefore likely to respond well to similar techniques. Both types of networks tend to be scale-free with low average diameter and high levels of clustering \cite{jackson2005economics}, and both have real-world meaning and are associated with significant amounts of external information. This real-world meaning makes alignment strategies an obvious choice; local alignment could be applied to community detection problems, and global alignment could be used to superimpose a user network from one source onto another in order to link user identities.

Finally, the use of the assignment problem in pattern recognition seems fairly niche despite being a dominant and well-explored strategy in biology, and computer scientists could likely find inspiration in global alignment techniques. They could also explore the idea of favoring connectivity in a mapping result; the seed-and-extend strategies used to construct well-connected alignments might be an effective strategy for quickly ruling out or narrowing down search areas for a potential subgraph isomorphism between a small graph and a much larger one.

\begin{table}[p]
\centering
\begin{tabular}{| l | l | l |}
\hline
\textbf{Computer Science} & \textbf{Biology} & \textbf{Mathematics} \\ \hline\hline
ACM & Biochem- & Algebra \\
Algorithm & Biocomputing & Algorithm \\
Artificial Intelligence & Bioengineering & Chaos \\
CIVR & Bioinformatic & Combinatori- \\
Computational Intelligence & Biological & Fixed Point \\
Computational Linguistics & Biology & Fractal \\
Computer & Biomedic- & Functional Analysis \\
Computer Graphics & Biosystem & Geometr- \\
Computer Science & Biotechnology & Graph \\
Computer Vision & Brain & Kernel \\
Data & Cancer & Linear Regression \\
Data Mining & Cardiology & Markov \\
Document Analysis & Cell & Mathemati- \\
Electrical Engineering & Disease & Multivariate \\
Graphics & DNA & Network \\
IEEE & Drug & Optimization \\
Image Analysis & Endocrinology & Permutation Group \\
Image Processing & Epidemiology & Probability \\
Intelligent System & Genetic & Riemann Surface \\
Internet & Genome & SIAM \\
ITiCSE & Genomic & Statistic- \\
Language Processing & Medical & Topology \\ 
Learning & Medicinal & Wavelet \\
Machine Learning & Medicine & \\
Machine Vision & Metabolic & \\
Malware & Microbiology & \\
Neural Network & Molecular & \\
Pattern Recognition & Neuro- & \\ 
Robotic & Neurobiological & \\
Scientific Computing & Pathology & \\
SIAM & Pathogen & \\
Signal Processing & Pharma- & \\
Software & Plant & \\
World Wide Web & Protein & \\
 & Proteom- & \\
 & Psych- & \\
 & Psychology & \\
 & Virology & \\
 & Virus & \\
\hline
\end{tabular}
\caption{Keywords used to tag journal names as various subjects.}
\vspace{-6pt}\flushleft\footnotesize *Note: Both a term and its plural are considered a match, and hyphens indicate a word with several ending variations which were all considered to be associated with the tag. While the search process was case sensitive in order to avoid false positives for short words like ``ACM", case-insensitive duplicate words have been excluded from the table. The words ``algorithm" and ``SIAM" are considered to be both computer science and mathematics.
\label{tab:tagging_keywords}
\end{table}
\bibliographystyle{plain}
\bibliography{thesis_bibliography}

\printindex

\end{document}